\def\Re{{\rm Re}}
\def\Im{{\rm Im}}
\def\PiL{\Pi_{\rm L}}
\def\PiT{\Pi_{\rm T}}
\def\tpp{\tilde {\bf p}_\perp}
\def\tqp{\tilde {\bf q}_\perp}
\def\tpsq{{\tilde{p}_{\perp}^2}}
\def\tqsq{{\tilde{q}_{\perp}^2}}
\def\tbf{\tilde{\bf f}}
\def\dEtw{{\widetilde{\delta E}\rule{0ex}{2.3ex}}}

\def\ppar{{p_\parallel}}
\def\qpar{{q_\parallel}}
\def\ipa{inelastic pair annihilation}
\def\Ctwo{C_{2\leftrightarrow 2}}
\def\LPM{{\scriptscriptstyle \rm LPM}}
\def\GammaE{\nu_{\rm e}}
\def\GammaA{\nu_{\rm a}}

\def\p{{\bf p}}
\def\q{{\bf q}}
\def\k{{\bf k}}
\def\f{{\bf f}}

\def\cf{C_{\rm F}}
\def\ta{T_{\rm A}}
\def\tf{T_{\rm F}}
\def\df{d_{\rm F}}
\def\nf{N_{\rm f}}
\def\nc{N_{\rm c}}
\def\mD{m_{\rm D}}
\def\mQ{m_{\rm Q}}
\def\Eg{k}
\newcommand\ansatz{{\it Ansatz}}

\newcommand\Eq[1]{Eq.~(\ref{#1})}

\def\half{{\textstyle{1\over2}}}
\def\gs{g_{\rm s}}
\def\alphas{\alpha_{\rm s}}
\def\alphaEM{\alpha_{\rm EM}}

\def\gsim{\mbox{~{\protect\raisebox{0.4ex}{$>$}}\hspace{-1.1em}
	{\protect\raisebox{-0.6ex}{$\sim$}}~}}
\def\lsim{\mbox{~{\protect\raisebox{0.4ex}{$<$}}\hspace{-1.1em}
	{\protect\raisebox{-0.6ex}{$\sim$}}~}}

\documentstyle[preprint,aps,fixes,epsf,eqsecnum]{revtex}

\makeatletter           
\@floatstrue
\def\figure{\let\@capwidth\columnwidth\@float{figure}}
\let\endfigure\end@float
\@namedef{figure*}{\let\@capwidth\textwidth\@dblfloat{figure}}
\@namedef{endfigure*}{\end@dblfloat}
\def\table{\let\@capwidth\columnwidth\@float{table}}
\let\endtable\end@float
\@namedef{table*}{\let\@capwidth\textwidth\@dblfloat{table}}
\@namedef{endtable*}{\end@dblfloat}
\def\la{\label}
\makeatother

\tightenlines
\newcommand\pcite[1]{\protect{\cite{#1}}}

\begin {document}


\preprint {UW/PT 01--22}

\title
    {
    Photon Emission from Quark-Gluon Plasma:\\
    Complete Leading Order Results
    }

\author {Peter Arnold}
\address
    {%
    Department of Physics,
    University of Virginia,
    Charlottesville, Virginia 22901
    }%
\author{Guy D. Moore and Laurence G. Yaffe}
\address
    {%
    Department of Physics,
    University of Washington,
    Seattle, Washington 98195
    }%

\date {November 2001}

\maketitle
\vskip -20pt

\begin {abstract}%
    {%
	We compute the photon emission rate of an equilibrated,
	hot QCD plasma at zero chemical potential,
	to leading order in both $\alphaEM$ and the QCD coupling
	$\gs(T)$.
	This requires inclusion of near-collinear bremsstrahlung
	and \ipa\ contributions,
	and correct incorporation of Landau-Pomeranchuk-Migdal
	suppression effects for these processes.
	Analogous results for a QED plasma are also included.
    }%
\end {abstract}

\thispagestyle{empty}

\section{Introduction}
\la{sec:intro}

How brightly does the quark-gluon plasma glow?  With the commissioning
of RHIC and the future commissioning of the LHC in heavy ion mode, this
is becoming an experimental question.  It seems appropriate, then, to
complete a theoretical analysis of photo-emission from a central heavy
ion collision at RHIC or LHC energies.

The goal of this paper is more modest.  We will be concerned only with
a fully equilibrated quark-gluon plasma, at a temperature high enough that
perturbation theory is applicable.%
\footnote
    {
    Since we take the plasma to be thermalized, the spatial size $R$
    of the plasma must exceed the mean free path for large angle
    scattering of a quark or gluon;
    this requires $\alphas^2 \, RT \gg 1$ at temperature $T$.
    }
In this context we will compute the
photo-emission rate to leading order in the strong coupling
constant $\gs$.
By this, we mean the spontaneous emission rate for photons of a given
momentum $\k$, which we will denote as
\begin {equation}
    \GammaE(\k) \equiv (2\pi)^3 \; {d \Gamma_\gamma \over d^3\k} \,.
\end {equation}
Our normalization is chosen so that
$\Gamma_\gamma \equiv \int \GammaE(\k) \> d^3\k/(2\pi)^3$
is the total spontaneous photon emission rate per unit volume.%
\footnote
    {
    If $f_\gamma(\k)$ denotes the phase space number density of photons,
    and $n_\gamma \equiv \int f_\gamma(\k) \; d^3\k/(2\pi)^3$
    is the total number of photons per unit volume,
    then the evolution of the photon phase space density is given by
    $
	{df_\gamma(\k) / dt}
	=
	\GammaE(\k) \, [1 + f_\gamma(\k)] - \GammaA(\k) \, f_\gamma(\k)
    $,
    where $\GammaE(\k)$ is the spontaneous emission rate
    and $\GammaA(\k)$ is the corresponding absorption rate.
    As always, detailed balance relates these rates,
    $\GammaA(\k) = \GammaE(\k) \, e^{|\k|/T}$.
    Photon absorption can be ignored if the plasma is optically thin;
    this requires that $R \, \GammaA \ll 1$.
    For a quark-gluon plasma at temperature $T$,
    this condition (as we will see) becomes
    $\alphas \, \alphaEM  \, R T \ll 1$.
    This should hold for all realistic heavy ion scenarios.
    In contrast, photon absorption can never be ignored in a thermalized
    relativistic QED plasma, since the photon absorption rate is
    comparable to the electron large-angle scattering rate.
    }

\begin{figure}[t]
\centerline{\epsfbox{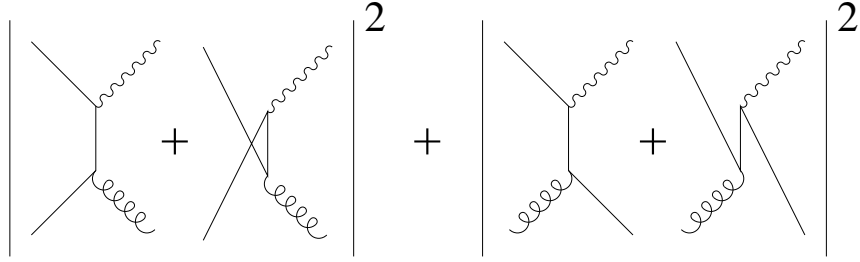}}
\vspace{0.1in}
\caption{\label{fig:2to2} Two-to-two particle processes which generate
the leading logarithmic contribution to the photo-production rate,
and were originally believed to give the complete leading order contribution.
Time may be viewed as running from left to right.}
\end{figure}

The evaluation of the photo-emission rate has quite a long history 
\cite{oldie1,oldie2,oldie3,oldie4,oldie5,oldie6,oldie7,oldie8},
and was thought to have been settled ten years ago when
Kapusta {\em et~al.}~\cite{Kapusta} and Baier {\em et~al.}~\cite{Baier}
independently computed the diagrams shown in Fig.~\ref{fig:2to2},
and obtained the same answer.
The result of Refs.~\cite{Kapusta,Baier},
written in a slightly more general fashion than given in those papers,
is
\begin{equation}
\label{eq:2to2}
    \GammaE(\k) =
    {\cal A}(k)
    \left[\,
	\ln ({T}/{m_\infty}) + \half \ln (2{\Eg}/{T}) + \Ctwo (\Eg/T) \,
    \right] ,
\end{equation}
where $\Eg \equiv |\k|$ is the energy of the photon.
The leading-log coefficient ${\cal A}(k)$ is given by
\begin {equation}
    {\cal A}(k)
    =
    2\, \alphaEM
    \Big[\, \df \sum_{s} q_{s}^2 \,\Big] \;
    \frac{m_\infty^2}{k} \>
    n_f(\Eg) \,
    ,
\label{eq:calA}
\end {equation}
where $n_f(\Eg) = [\exp(\Eg/T)+1]^{-1}$ is the Fermi distribution function,
and $\df$ is the dimension of the quark representation
(so $\df=\nc=3$ for QCD).
The charge assignment for each quark species $s$ is denoted
$q_{s}$, and equals 2/3 for up type and $-1/3$ for down type quarks.
The mass $m_\infty$ appearing in
Eqs.~(\ref {eq:2to2}) and (\ref {eq:calA})
is the leading-order asymptotic (large momentum) thermal quark mass,%
\footnote
    {
    The momentum-dependent thermal mass $m(p)$ is defined by the
    relation $E(p) = \sqrt {p^2 + m(p)^2}$, where $E(p)$ is
    the energy of a ``dressed'' quark moving through the plasma
    with spatial momentum $p$ (and whose zero-temperature mass
    is negligible compared to $\gs \, T$).
    For a typical momentum of order $T$, one may expand the
    square root and see that the thermal correction to the
    quark dispersion relation is proportional to
    the thermal mass squared,
    $\Delta E(p) \equiv E(p) - p \sim m(p)^2 / (2p)$.
    For $p = O(T)$ or larger, the momentum-dependent thermal mass
    equals the the asymptotic mass $m_\infty$,
    up to irrelevant sub-leading corrections suppressed by additional
    powers of $\gs$.
    This asymptotic value of $m(p)$ is related to the thermal mass $\mQ$,
    conventionally defined as the energy of a dressed quark which is
    at rest in the plasma, by a square root of two,
    $m_\infty^2 = 2 \mQ^2$.
    }
given by \cite{Weldon}
\begin{equation}
    m_\infty^2 = {\cf \, \gs^2 \, T^2 \over 4} \,,
\label {eq:minfty}
\end{equation}
with $\cf$ the quadratic Casimir of the quark representation.
For the fundamental representation of SU($\nc$),
$\cf = (\nc^2{-}1)/(2\nc)$;
for QCD specifically, $\cf = 4/3$ and $m_\infty^2 = {\gs^2 \, T^2}/{3}$.
Hence,
for two-flavor QCD, the leading-log coefficient is explicitly
\begin {equation}
    {\cal A}(k)
    =
    {40 \, \pi T^2 \over 9} \, \alphaEM \, \alphas \,
    {n_f(\Eg) \over k} \, .
\end {equation}
The strong coupling $\gs$ (and $\alphas \equiv \gs^2/4\pi$) should always
be understood as defined at a scale of order of the temperature.%
\footnote
    {%
    For QED, $\cf = 1$ and $m_\infty^2 = {e^2 \, T^2}/{4}$.
    Hence
    ${\cal A}(k)$ equals $2\pi \, \alphaEM^2 \, n_f(\Eg)/k$
    for an electromagnetic
    plasma containing only electrons and positrons.
    }

The final term $\Ctwo(\Eg/T)$ in expression (\ref {eq:2to2})
is a function of $\Eg/T$ with a finite limit as $\Eg/T \to \infty$.
The papers \cite {Kapusta,Baier} compute $\Ctwo(\Eg/T)$ only in this limit,
and find
\begin{equation}
\label{eq:K_is}
\lim_{\Eg/T \rightarrow \infty} \Ctwo(\Eg/T) = 
	-\frac{1}{4} - \frac{\gamma_{\rm E}}{2} + \frac{2\ln(2)}{3}
	+ \frac{\zeta'(2)}{2 \zeta(2)} \simeq -0.3614902 \, .
\end{equation}
We have also verified this result.

\begin{figure}[t]
\centerline{\epsfbox{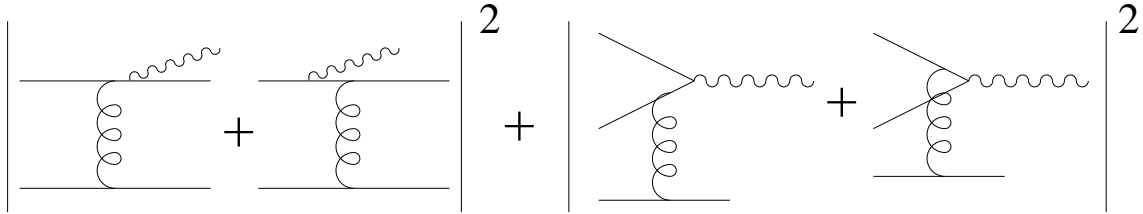}}
\vspace{0.1in}
\caption{\label{fig:inelastic}
Inelastic processes which, because of near-collinear singularities,
contribute at the same order as the two-to-two particle processes.
The diagrams on the left represent bremsstrahlung, and those on the right
are \ipa.  Again, time may be viewed as running from left to right.}
\end{figure}

However, as recently demonstrated by Aurenche
{\em et~al.}~\cite{Gelis1,Gelis2,Gelis3}, the result (\ref {eq:2to2})
is incomplete.
The bremsstrahlung and \ipa\ processes shown in Fig.~\ref{fig:inelastic}
contain collinear enhancements which cause them to contribute at
the same parametric order in coupling, $O(\alphaEM \, \alphas)$,
as the two-to-two processes of Fig.~\ref{fig:2to2},
even for large photon energies $\Eg \gg T$.
But the calculation of these processes presented
in \cite{Gelis1} is also incomplete,%
\footnote{It is also wrong by a factor of 4 \pcite{SteffenThoma}.} 
as it fails to incorporate an $O(1)$ suppression
of these processes due to
multiple scattering during the photon emission process,
which limits the coherence length of the emitted radiation.
This is known as the Landau-Pomeranchuk-Migdal (LPM) effect
\cite{LP,M1,M2}.
The presence of LPM suppression was pointed out in later work by
Aurenche {\em et~al.}~\cite{Gelis2,Gelis3},
which discussed the physics involved but did not attempt to make a complete
calculation.

This paper presents a full calculation of the photon emission rate at
leading order in $\gs$.
We evaluate the non-logarithmic two-to-two contribution $\Ctwo(\Eg/T)$
for general $\Eg/T$.
We also compute the rate of photo-production by bremsstrahlung and
\ipa, fully including the LPM effect.
This requires solving a non-trivial integral equation to determine this rate.
We derived this integral equation in Ref.~\cite{us};
it is related to one written down over 45 years ago
by Migdal~\cite{M1,M2} for studying electromagnetic energy loss
of a relativistic particle traversing matter.
Related work in the QCD context may be found in
Refs.~\cite{Zakharov,BDMPS,LPM_QCD1,LPM_QCD2,LPM_QCD3,LPM_QCD4}.%
\footnote
   {%
    References \cite{Zakharov} and \cite{BDMPS},
    which predate our work \cite{us},
    derive an integral equation similar to our result.
    The main difference is that we treat the actual distribution
    of moving colored charges in the plasma, with dynamical screening,
    rather than employing a model of static colored scattering centers
    as in \cite{Zakharov,BDMPS}.
    Ref.~\cite{us} also includes plasma induced corrections to
    quasiparticle dispersion relations (which do influence
    the leading-order result), and contains a fairly elaborate
    diagrammatic power counting analysis to convincingly identify
    all leading order effects.
   }
Other discussions of the LPM effect for
electromagnetic interactions in ordinary matter include
Refs.~\cite {LPM_QED1,LPM_QED2,LPM_QED3}.
The current paper is a sequel to our earlier work \cite{us},
but it should be fairly self-contained.

The outline of the remainder of this paper is as follows.
In Sec.~\ref{sec:review}, we briefly
review our previous work \cite{us} showing why the processes
considered here (and no others) contribute to the leading order emission rate,
and deriving an integral equation whose solution determines
the bremsstrahlung and \ipa\ contributions to the emission rate,
fully including the LPM effect.
Sec.~\ref{sec:lpm} discusses, qualitatively, the relative
importance of the LPM effect in different kinematic regimes.
The numerical solution of the integral equation
determining the LPM contributions to the emission rate
is presented in Sec.~\ref{sec:brem}.
This section is the meat of the paper.
Sec.~\ref{sec:K} contains the evaluation of $\Ctwo(\Eg/T)$
for arbitrary photon energy $\Eg \gg \gs T$.
(If this condition is violated, then thermal corrections
to the dispersion relations for incoming or outgoing particles
can no longer be neglected in these $2\leftrightarrow2$ processes.)
Our results are combined in
Sec.~\ref{sec:results}, which is followed by a brief conclusion.

For the convenience of readers interested in just the bottom line,
we summarize our results here.
The complete leading-order photon emission rate may be written as
\begin{eqnarray}
    \GammaE(\k) &=&
    {\cal A}(k) \,
    \Bigl[ \, \ln \left({T}/{m_\infty}\right) + C_{\rm tot}(k/T) \, \Bigr] \,,
\label {eq:preview_rate}
\\
\noalign {\hbox{with}}
    C_{\rm tot}(k/T) &\equiv& \half \ln \left({2k}/{T}\right)
    + \Ctwo(k/T) + C_{\rm brem}(k/T) + C_{\rm annih}(k/T) \,,
\label {eq:preview_Ctot}
\end{eqnarray}
and the coefficient ${\cal A}(k)$ given in \Eq{eq:calA}.
The function $C_{\rm tot}(k/T)$ is the total ``constant under the log'';
it is a non-trivial function of $k/T$
but it is independent of the strong coupling $\gs$, whereas
$\ln (T/m_\infty) \sim \ln (1/\gs)$ since $m_\infty = O(\gs \, T)$.
Corrections to \Eq{eq:preview_rate} are suppressed by one or more powers
of $\gs$.

The functions $\Ctwo(k/T)$, $C_{\rm brem}(k/T)$, and $C_{\rm annih}(k/T)$
all involve multidimensional integrals (or integral equations) which
cannot be evaluated analytically but may be computed numerically.
Individual plots of these functions, as well as their sum,
are shown in section~\ref {sec:results}.
Our numerical results for QCD plasmas are reproduced quite accurately
by the approximate, phenomenological fits
\begin {eqnarray}
    \Ctwo(x)
    &\simeq&
    0.041 \, x^{-1} - 0.3615 + 1.01 \, e^{-1.35 \, x} \,,
\label {eq:approxC2}
\\ \noalign{\hbox{and}}
    C_{\rm brem}(x) + C_{\rm annih}(x)
    &\simeq&
      \sqrt{1 {+} {\textstyle {1\over6}} \nf}
      \left[
	\frac{0.548 \, \log(12.28 + 1/x)}{x^{3/2}}
	+ \frac{0.133 \, x}{\sqrt{1+x/16.27}} 
    \right] ,
\label {eq:approxC}
\end {eqnarray}
where $\nf$ is the number of quark flavors.
The approximation (\ref {eq:approxC2}) to the function $\Ctwo(x)$
(which crosses zero) is accurate to within an absolute error of 0.02
over the range $0.2 \le k/T \le 50$, but this form fails to be a good
approximation at smaller $k/T$;
$\Ctwo(x)$ has a finite small $x$ limit, rather than growing as $1/x$.
As discussed in section \ref {sec:lpm}, the form (\ref {eq:approxC})
for the near-collinear contributions builds in the right parametric
large and small $k$ asymptotic behavior, up to logs,
and is accurate to within a 3\% relative error over the range
$0.2 \le k/T \le 50$ for $\nf$ from 2 to 6.%
\footnote
    {
    It should be emphasized that
    throughout our analysis \cite{us} we assume that the photon
    wavelength is much smaller than the large angle mean free path for quarks.
    This requires that the photon energy be
    parametrically large compared to $\gs^4 T\ln \gs^{-1}$.
    We also require $k \gg m_\gamma$, where $m_\gamma$ is the
    asymptotic (electromagnetic) thermal mass of the photon,
    given in Eq.~(\protect\ref{eq:mgamma}).
    This latter condition ensures that plasma corrections
    to the photon dispersion relation are small and do not
    significantly reduce the photon velocity below $c$.
    Given the physical value of $\alphaEM$, this implies
    that our results for soft bremsstrahlung can only be trusted down to
    about $k \approx 0.2 \, T$ (for QED with only electrons and positrons)
    or $k \approx 0.3 \, T$ (for three flavor QCD).
    }

Consider, for the sake of discussion, a value of $\alphas = 0.2$
which corresponds to $\ln (T/m_\infty) = 0.09$.
In this case,
it turns out that the old results of Baier {\em et~al.}~\cite{Baier}
and Kapusta {\em et~al.}~\cite{Kapusta},
which neglect near-collinear bremsstrahlung and \ipa,
are within a factor of two of the correct 
leading-order rate
for photon momenta in the range $2.5 \le k/T \le 10$.
Bremsstrahlung becomes the most important process for energies below this range,
and \ipa\ makes a large relative contribution to the rate above this
range (where the total rate is small because of exponentially falling
population functions).
The LPM effect suppresses both bremsstrahlung
and pair annihilation processes, but the suppression is not severe
(35\% or less) except for soft bremsstrahlung with $\Eg \le 2T$,
or very hard pair annihilation with $\Eg \gg 10T$,
where the LPM suppression is significant.
For parametrically small photon energy
$\Eg \ll T$ (but with $k \gg eT$ and $k \gg \gs^4 T \ln \gs^{-1}$), 
the LPM effect changes the parametric behavior of the emission rate.
One finds,
\begin{equation}
{d\Gamma_\gamma} \propto \Eg^{-1/2} \, d\Eg \, ,
\end{equation}
[ignoring $\log(T/k)$ factors],
rather than $\Eg^{-1} \, d\Eg$ which is the
result if one ignores the LPM effect.
Therefore some results in the literature \cite{Gelis1,softgamma}
concerning soft, on-shell photo-emission from the quark gluon plasma,
ignoring the LPM effect, are wrong {\em parametrically}.
Similarly, for extremely hard photon emission, the rate behaves as
$\Eg^{3/2} \, e^{-\Eg/T} dk$, rather than $\Eg^2 \, e^{-\Eg/T} dk$
as one would obtain by neglecting the LPM effect.
For more detailed results see the figures in Sec.~\ref{sec:results}.

\section{Bremsstrahlung and inelastic pair annihilation}
\label{sec:review}

\begin{figure}
\centerline{\epsfbox{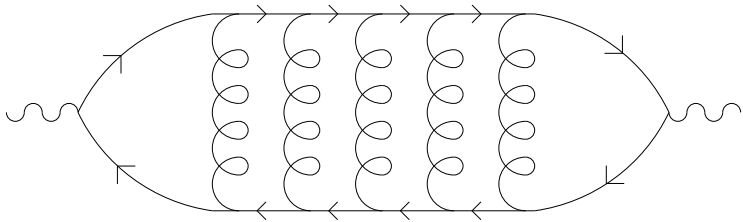}}
\vspace{0.1in}
\caption{Typical ladder diagram contributing to the electromagnetic
current-current correlator.
All such ladder diagrams must be summed to determine the leading-order
bremsstrahlung and \ipa\ rates.
Resummation of self-energy insertions on all propagators is implied.
\label{fig:brem_pic}}
\end{figure}

\begin{figure}
\centerline{\epsfxsize=6.3in\epsfbox{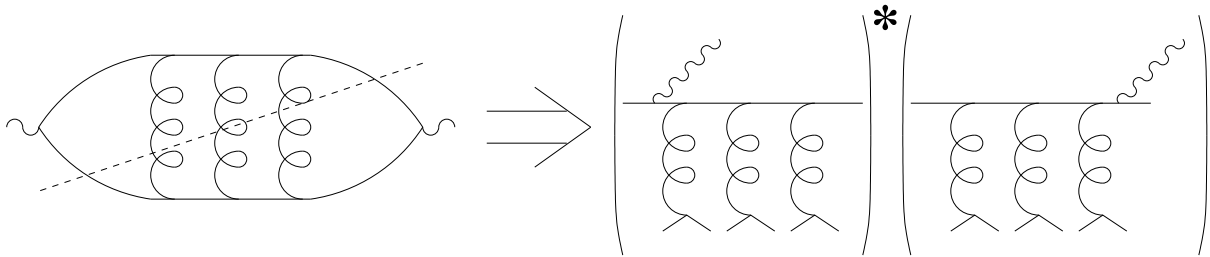}}
\vspace{0.1in}
\caption{Cut ladder diagram with $N$ gluon ``rungs'' may be interpreted
as the interference between amplitudes for photon emission before and
after $N$ scattering events.
\label{fig:interp}}
\end{figure}

The photo-emission rate is related to the imaginary part of
the electromagnetic current-current correlator.
We show in \cite{us} that a correct treatment of the near-collinear
bremsstrahlung and \ipa\ processes which contribute to the
leading-order emission rate requires a summation of all
ladder graphs of the form shown in Fig.~\ref{fig:brem_pic}.
(Self-energy insertions on all propagators,
including the leading order imaginary parts,
are implicitly resummed.)
As indicated in Fig.~\ref{fig:interp},
these diagrams can be interpreted as generating
various contributions to the squares of scattering amplitudes.
It is necessary to sum all ladders because there can
be interference between emission at one scattering and at subsequent
scattering events.
This is the basic origin of the LPM effect.

The total photon emission rate is given by an integral over all particle
momenta of the emission rate for that particle momentum.  Factoring out
the same coefficient ${\cal A}(k)$
which appears in the leading logarithmic result (\ref {eq:2to2}),
we find that the contribution of bremsstrahlung and \ipa\ processes
to the leading-order emission rate is \cite{us}
\begin{eqnarray}
    \GammaE^\LPM(\k) & = &
	{\cal A}(k)
	\!\int_{-\infty}^{\infty} \!\! d\ppar
	\left[
	    \frac{\ppar^2+(\ppar{+}k)^2} {4\ppar^2 \, (\ppar{+}k)^2}
	\right] \!
	\frac{n_f(k{+}\ppar)\, [1{-}n_f(\ppar)]}{m_\infty^2 \, n_f(k)}
	\! \int \! \frac{d^2 \p_\perp}{(2\pi)^2} \,
	\Re \left\{ 2 \p_\perp \! \cdot \f(\p_\perp,\ppar,k) \right\} \! .
\nonumber\\
\label{eq:outer_int}
\end{eqnarray}
Here $\ppar$ and $\p_\perp$ are the components of the momentum $\p$
parallel or perpendicular to the photon direction, respectively.
The contribution from the integration region $-k<\ppar<0$,
in which $\ppar$ and $(\ppar{+}k)$
have opposite sign and $1{-}n_f(\ppar) = n_f(|\ppar|)$, should be
interpreted as \ipa, while the contributions from $\ppar>0$ or
$\ppar<-k$ represent bremsstrahlung off a particle of energy $\ppar{+}k$
(when $\ppar>0$) or an anti-particle of energy $-\ppar$ (for $\ppar<-k$). 

The final term $\Re \{ 2 \p_\perp \cdot \f(\p_\perp,\ppar,k) \}$ 
is proportional to the amplitude squared for a particle of momentum
$\p{+}\k$ to emit a photon of momentum $\k$,
and is the non-trivial part of the calculation.
The function $\f(\p_\perp, \ppar , k)$ is the solution to
an integral equation, derived in \cite{us},
which accounts for the partial coherence between multiple scattering events.
This integral equation is
\begin{eqnarray}
\label{eq:f_with_dims}
    2\p_\perp & = & i \, \delta E(\p_\perp) \: \f(\p_\perp)
    +
    \int {d^2\q_\perp \over (2\pi)^2} \; {\cal C}(\q_\perp) \,
    \Big[ \f(\p_\perp) -  \f(\p_\perp{+}\q_\perp) \Big] \,,
\end{eqnarray}
where the dependence of $\f(\p_\perp,\ppar,k)$ (as well as $\delta E$)
on $\ppar$ and $k$ has been suppressed for notational convenience.
The collision kernel ${\cal C}(\q_\perp)$ is given by
\begin {equation}
    {\cal C}(\q_\perp)
    \equiv
    \frac{\mD^2 \, m_\infty^2}{T^2}
	\int {dq_\parallel \, dq^0}
	\: \delta(q_\parallel{-}q^0) \,
	 \frac{T}{q}
	\left\{
	    \frac{2}{\left|q^2 - \PiL(q^0,q)\right|^2}
	    +
	    \frac{[1-(q^0/q)^2]^2}
		 {\left|(q^0)^2-q^2-\PiT(q^0,q) \right|^2}
	\right\} ,
\label {eq:Cdef}
\end{equation}
where $q \equiv |\q| = [\q_\perp^2 + \qpar^2]^{1/2}$,
$\mD$ is the (leading order) Debye mass,
and $\PiL(q^0,q)$ and $\PiT(q^0,q)$ are the standard hard thermal loop
longitudinal and transverse gluon self-energies \cite {Weldon1},
respectively; their explicit form is given in Section \ref{sec:brem}.
In the first term on the right-hand side of \Eq{eq:f_with_dims},
$\delta E$ denotes the energy difference, on shell,
between the two relevant states of the system, in the limit where $\ppar$
is large compared to $|\p_\perp|$ and $m_\infty$.
For bremsstrahlung, this is the energy difference
between a quark of momentum $\p$ plus a photon of momentum $\k$,
and a quark of momentum $\p{+}\k$, which explicitly is
\begin {eqnarray}
    \delta E(\p_\perp,\ppar,k)
    &=&
    \left( \ppar + {\p_\perp^2 + m_\infty^2 \over 2\ppar} \right)
    +
    \left( k + {m_\gamma^2 \over 2k} \right)
    -
    \left( \ppar + k + {\p_\perp^2 + m_\infty^2 \over 2(\ppar{+}k)} \right)
\nonumber\\ &=&
	\frac{k}{2\ppar(k{+}\ppar)} 
	\left[ \p_\perp^2 + m_\infty^2 \right] + \frac{m_{\gamma}^2}{2k} \,.
\label{eq:dE_with_dims}
\end {eqnarray}
For pair annihilation, $\delta E$ is the difference between a quark plus
anti-quark with momenta $-\p$ and $\p{+}\k$, and a photon of momentum $\k$;
this gives exactly the same result (\ref {eq:dE_with_dims})
given our convention about the sign of $\ppar$.
This energy difference generates a phase difference
between the amplitudes in Fig.~\ref{fig:interp},
which is manifested as an imaginary term in \Eq{eq:f_with_dims}.
This phase difference will eliminate coherence in the emitted
radiation beyond a time scale of order $1/\delta E$,
which is often referred to as the formation time of the photon.
The $m_{\gamma}^2$ contribution to $\delta E$ accounts
for plasma induced dispersion in the photon propagation;
$m_\gamma$ is the asymptotic thermal photon mass,
whose square is half the {\em electromagnetic} Debye mass
squared of the plasma,
\begin {equation}
    m_\gamma^2
    = \half \, m_{{\rm D},\gamma}^2
    = {1\over 6} \, \Big[ \, \df \sum_{s} q_{s}^2 \, \Big] \,
	e^2 \, T^2 \,.
\label {eq:mgamma}
\end {equation}
Since we assume $\alphaEM \ll \alphas$, we may neglect this
electromagnetic dispersion correction when considering photo-emission
from a quark-gluon plasma;
but when we consider the emissivity of an electromagnetic plasma,
which we will also compute, retaining this term will be essential. 

Note that the collision kernel (\ref {eq:Cdef}) is rotationally
invariant (in the transverse plane).
Consequently, the solution to the integral equation (\ref {eq:f_with_dims})
must equal $\p_\perp$ times some scalar function of $\p_\perp^2$,
\begin {equation}
    \f(\p_\perp) = \p_\perp \, \chi(p_\perp^2) \,.
\label {eq:rot_inv}
\end {equation}
Hence, although written as a vector equation,
the integral equation (\ref {eq:f_with_dims}) is trivially reducible to
a scalar integral equation.
For later convenience, also note that
this integral equation may be written more abstractly as
$
    2\p_\perp = [ i \, \delta E + \hat {\cal C} \, ] \, \f
$,
if one defines $\hat {\cal C}$ as a linear operator
acting on functions of $\p_\perp$, whose explicit form
is given by the second term in \Eq{eq:f_with_dims}.
Further discussion of equation (\ref {eq:f_with_dims}),
and its derivation, may be found in Ref.~\cite{us}.  

\section {LPM effect: Rough Estimates}
\label {sec:lpm}

Before solving equation (\ref {eq:f_with_dims}) numerically,
it is useful to examine in what kinematic regimes
the LPM effect will be important
and to estimate how large its effect can be.
Neglecting the LPM effect altogether means ignoring any interference
between successive scattering processes.
This corresponds to solving \Eq{eq:f_with_dims} under the
assumption that $\delta E$ is much greater than the collision
term which follows it.%
\footnote
    {%
    Note that $\delta E$ does not vanish for any value of
    $\p_\perp$ and $\ppar$ provided the inequality
    $2 m_\infty > m_\gamma$ is satisfied.
    This holds in QCD because of our assumption that $e$ is
    much smaller than $\gs$.  It is also satisfied for electron
    and electron plus muon QED plasmas.
    If this inequality is violated, then a photon propagating through
    the plasma could decay into a fermion-antifermion pair without
    any associated inelastic scattering, and the resulting pole
    in (\ref {eq:fguess}) would produce a real contribution to
    $\f(\p_\perp)$ proportional to
    $\p_\perp \, \delta (\delta E(\p_\perp))$.
    }
At leading order in this ``approximation,''
$\f(\p_\perp)$ is then pure imaginary,
\begin{equation}
    \Im \, \f(\p_\perp)\Big|_{\rm no \; LPM}
    =
    -\frac{2\p_\perp}{\delta E(\p_\perp)} \, .
\label {eq:fguess}
\end{equation}
The next order arises from substituting this result into the 
collision term, thereby determining the leading real part of $\f$
in this approximation,
\begin{equation}
    \Re \, \f(\p_\perp)\Big|_{\rm no \; LPM} 
    = \frac{1}{\delta E(\p_\perp)} 
    \int {d^2\q_\perp \over (2\pi)^2} \; {\cal C}(\q_\perp)
    \left[
	\frac{2\p_\perp}{\delta E(\p_\perp)} - 
	\frac{2(\p_\perp{+}\q_\perp)}{\delta E(\p_\perp{+}\q_\perp)}
    \right]
    \, .
\end{equation}
Inserting this form for $\Re \, \f(\p_\perp)$ into \Eq{eq:outer_int}
reproduces the results of Aurenche {\em et~al.}~\cite{Gelis1}
(except for the overall factor of 4 error made there \cite{SteffenThoma}).
Given the form (\ref {eq:dE_with_dims}) of $\delta E$,
and neglecting $m_\gamma^2$,
one may see without performing any calculation that 
$\Re \, \f(\p_\perp)|_{\rm no \; LPM}$ will be a $\ppar$ and $k$
independent function times $\ppar^2(\ppar{+}k)^2/k^2$.
Therefore, in this approximation, the photon spectrum will look like
\begin{equation}
    \GammaE(\k)\Big|_{\rm no \; LPM}
    \propto {\cal A}(k)
	\int_{-k/2}^\infty d\ppar
	\left[ \frac{\ppar^2 + (\ppar{+}k)^2}{k^2 \, T} \right] 
	\frac{n_f(k{+}\ppar)\, [1{-}n_f(\ppar)]}
	{n_f(k)} \, ,
\end{equation}
with a $k$-independent constant of proportionality.
For soft photons ($k\ll T$) the dominant contribution comes from
$\ppar \sim T$, and the result, ignoring the LPM effect, 
grows as $T^2/k^2$ relative to the emission rate due to only
$2\leftrightarrow2$ processes.
At large $k \gg T$ the
dominant contribution comes from the whole range $-k/2 < \ppar < 0$
over which the population function terms are approximately unity.
This yields a result (again, ignoring the LPM effect) 
which grows as $k/T$ relative to the $2\leftrightarrow2$ emission rate.
Therefore, ignoring the LPM effect, we
find that bremsstrahlung dominates the emission rate for $k \ll T$,
while \ipa\ dominates the rate for $k\gg T$.

We must now consider whether, for any range of $k$,
it is actually permissible to neglect the LPM effect.
Using the form (\ref {eq:dE_with_dims}) of $\delta E$,
one may easily see that if $kT / |\ppar \, (\ppar{+}k)|$
is sufficiently large then $|\delta E|$ will also be large for
all $\p_\perp$ and an expansion based on $|\delta E| \gg \hat{\cal C}$
will be valid for all $\p_\perp$.
However, in this regime, $\Re\, \f$ is always small,
and so is the photo-emission rate.
In both cases where we have just found a large photo-emission rate
(ignoring LPM suppression),
the quantity $kT / [\ppar \, (\ppar{+}k)]$ is {\em small}, and
hence the inequality
$\delta E \gg \hat{\cal C}$ only holds for large $\p_\perp$.
Therefore, the estimates we have just made are too optimistic,
and there will be substantial LPM suppression in exactly those regimes
where the inelastic rate is dominant.
To estimate how large the suppression is, we may solve for
$\f(\p_\perp)$ at large $\p_\perp$,
where the large $\delta E$ approximation is valid,
and determine where the approximation breaks down.
At large $\p_\perp$, we have
\begin{equation}
\label{eq:im_f1}
    \left. \Im \, \f(\p_\perp) \right|_{\rm asymptotic}
    = -\frac{2 \p_\perp}{\delta E(\p_\perp)}
    \simeq 
	-\left[\frac{2\ppar \, (\ppar{+}k)}{k} \right]
	\frac{2\p_\perp}{\p_\perp^2} \, ,
\end{equation}
and 
\begin{eqnarray}
\label{eq:re_f1}
    \left.\Re \, \f(\p_\perp)\right|_{\rm asymptotic} 
    &=&
    \frac{-1}{\delta E(\p_\perp)}\;
    \hat{\cal C}\: \Im \, \f|_{\rm asymptotic}
\nonumber
\\ &\simeq&
    \left[\frac{2\ppar \, (\ppar{+}k)}{k} \right]^2
    \frac{1}{\p_\perp^2} 
    \int {d^2\q_\perp \over (2\pi)^2} \; {\cal C}(\q_\perp)
    \left[
	\frac{2\p_\perp}{\p_\perp^2} - 
	\frac{2(\p_\perp{+}\q_\perp)}{(\p_\perp{+}\q_\perp)^2}
    \right]
    .
\label {eq:Refasym}
\end{eqnarray}
This integral may be evaluated in closed form.
Since the kernel ${\cal C}(\q_\perp)$ is rotationally invariant,
the integral in \Eq{eq:Refasym} may be rewritten as
\begin {equation}
    \int_0^\infty
    {d|\q_\perp| \over 2\pi} \> |\q_\perp| \>
    {\cal C}(|\q_\perp|)
    \left[
	{2\p_\perp \over \p_\perp^2}
	-
	\left< {2\p_\perp \over \p_\perp^2} \right>_{\!q_\perp}
    \right],
\label {eq:harm}
\end {equation}
where $\left< g(\p_\perp) \right>_{q_\perp}$ denotes the
average of a function $g$ over the circle of radius
$q_\perp \equiv |\q_\perp|$ centered at $\p_\perp$.
Next, note that (both components of) $\p_\perp/\p_\perp^2$ are
harmonic functions (away from $\p_\perp = 0$),
and recall that the average of a harmonic function around a circle
equals the value of the function at the center of the circle,
provided no singularities are enclosed.
Consequently, the integrand of \Eq{eq:harm} vanishes
for $|\q_\perp| < |\p_\perp|$.
For $|\q_\perp| > |\p_\perp|$,
elementary complex analysis shows that
$\left<\p_\perp/\p_\perp^2 \right>_{q_\perp}$ vanishes.%
\footnote
    {%
    If one relabels $\p_\perp$ as $\bf r$, and $q_\perp$ as $R$,
    then the reasoning just given is easily seen to be just a
    two dimensional version of Newton's classic result
    that the gravitational force due to a uniform spherical shell
    of radius $R$ vanishes inside the shell,
    but outside the shell is the same as the force due to a point mass
    at the center of the shell.
    }
Therefore,
\begin{equation}
    \left.\Re \, \f(\p_\perp)\right|_{\rm asymptotic} 
    =
    {1 \over \pi}
    \left[\frac{2\ppar \, (\ppar{+}k)}{k} \right]^2
    \frac{\p_\perp}{|\p_\perp|^4}
    \int_{|\p_\perp|}^\infty
    {\cal C}(|\q_\perp|) \, |\q_\perp| \, {d|\q_\perp|} \,.
\end{equation}
Since $\p_\perp$ has been assumed to be asymptotically large, and
$|\q_\perp| > |\p_\perp|$,
one may completely neglect the self-energy corrections in
the explicit form (\ref {eq:Cdef}) of ${\cal C}(\q_\perp)$.
The resulting integral is then straightforward, leading to
\begin{equation}
\label{eq:re_f2}
    \Re \, \f(\p_\perp)|_{\rm asymptotic} 
    =
    \left[ \frac{2\ppar(\ppar{+}k)}{k} \right]^2
    {\mD^2 \, m_\infty^2 \over \pi \, T} \,
    \frac{2\p_\perp}{|\p_\perp|^6} \, .
\end{equation}

These large $\p_\perp$ asymptotic forms are valid when
$\delta E \gg \hat{\cal C}$, or equivalently when
$
    \Re \, \f(\p_\perp)|_{\rm asymptotic} \ll
    \Im \, \f(\p_\perp)|_{\rm asymptotic}
$.
The expansion will have broken down when the real and imaginary parts
become comparable.
This occurs for
\begin{equation}
    \p_\perp^2
    \sim
    \left. \p_\perp^2 \right|_{\rm crossover}
    \equiv
    {\mD \, m_\infty} \,
    \sqrt{ \frac{\ppar\, (\ppar{+}k)}{kT}} \, .
\end{equation}
Below this crossover point,
$\Re\, \f(\p_\perp)$ will be smaller than the asymptotic estimate.
The region $\p_\perp \sim \p_\perp|_{\rm crossover}$ will, in fact,
provide the dominant contribution to the final $d^2\p_\perp$ integral
in the rate (\ref {eq:outer_int}).
Hence the size of this integral [ignoring factors of $\ln (k/T)$]
may be estimated as
\begin{eqnarray}
    \int_{\p_\perp^2|_{\rm crossover}}^\infty 
    d(p_\perp^2)  \>
    \Re \, \f|_{\rm asymptotic} \cdot \p_\perp 
    & \sim &
    \left[ \frac{\ppar \,(\ppar{+}k)}{k} \right]^2
    \frac{\mD^2 \, m_\infty^2}{T \, \p_\perp^2|_{\rm crossover}}
\nonumber \\ & \sim &
    \left[ \frac{\ppar(\ppar{+}k)}{kT} \right]^{3/2}
    {\mD \, m_\infty \, T}
    \, .
\end{eqnarray}

Using this result, we find that, for bremsstrahlung, the small $k$
emission rate grows, relative to the $2\leftrightarrow2$ rate,
only as $(T/k)^{3/2}$,
not $(T/k)^2$ as was found ignoring the LPM effect.
Similarly, for large $k$ the \ipa\ rate grows,
relative to the $2\leftrightarrow2$ rate, only as $(k/T)^{1/2}$,
not $(k/T)$ as was found when ignoring LPM.
The dependence on $\mD$ is also simple;
relative to the $2\leftrightarrow2$ rate,
the inelastic contribution grows linearly in the Debye mass $\mD$.
Therefore,
inelastic processes become relatively more important as we increase the
number of species in the plasma (which increases $\mD$),
but only as the square root of the number of species.

\section{Solving the integral equation}
\label{sec:brem}

In order to solve the integral equation (\ref {eq:f_with_dims}) which
determines $\f(\p_\perp)$, we will use an adaptation of the
variational approach we previously exploited in \cite{lead_log,largeN}.
But first, it will be convenient to scale out the dimensionful
parameters from \Eq{eq:f_with_dims}.
To do so, we define dimensionless transverse momenta,
\begin {equation}
    \tpp \equiv \frac{\p_\perp}{\mD} \, , \quad
    \tilde{\q} \equiv \frac{\q}{\mD} \, , \quad
\end {equation}
and correspondingly define
\begin {equation}
\dEtw(\tpp) \equiv \frac{T}{\mD^2} \, \delta E(\p_\perp) \,, \qquad
\tbf(\tpp) \equiv \frac{\mD}{T} \, \f(\p_\perp) \, .
\end{equation}
It will also be convenient to define the dimensionless ratio
of thermal masses
\begin{equation}
    \kappa \equiv \frac{m_\infty^2}{\mD^2}
    = \frac{3 \cf}{4(\ta + \nf \, \tf)}
    = \left\{ \begin{array}{ll}
	3/(4\nf) \, , & {\rm QED} \, ; \\
	2/(6+\nf) \, , \qquad & {\rm QCD} \, . \\
	\end{array} \right.
\end{equation}
Here $\ta$ and $\tf$ are the trace normalizations of the adjoint and
quark representations, respectively.
For QCD, $\ta = \nc=3$ and $\tf = 1/2$,
while for QED, $\ta = 0$, and $\tf = 1$.

Substituting these dimensionless variables into \Eq{eq:outer_int}, and
noting that the argument of the $\ppar$ integral 
is invariant under $\ppar \rightarrow -k-\ppar$, that equation becomes
\begin{eqnarray}
    \GammaE^\LPM(\k)
    & = &
	{{\cal A}(k) \, T \over \kappa}
	\int_{-k/2}^{\infty} d\ppar
	\left[
	    \frac{\ppar^2+(\ppar{+}k)^2} {2\ppar^2 \, (\ppar{+}k)^2}
	\right]
	\frac{n_f(k{+}\ppar)\, [1{-}n_f(\ppar)]}{n_f(k)} \>
	\Big( 2\tpp \, , \, \Re \, \tbf \Big) \, ,
\label{eq:outer}
\end{eqnarray}
where we have introduced a natural inner product on the
space of (vector valued) functions of transverse momentum,
\begin{equation}
\label{eq:inner_prod}
\Big( \f_1 \, , \, \f_2 \Big) \equiv \int 
	\frac{d^2 \tpp}{(2\pi)^2} \;
	\f_1(\tpp) \cdot \f_2(\tpp) \, .
\end{equation}
Once again, the contribution from $\ppar <0$ is \ipa,
while the contribution from $\ppar > 0$ is bremsstrahlung.
The dimensionless form of the integral equation for $\tbf$ is now
\begin{equation}
\label{eq:f_dimless}
    2 \tpp = \left[ i \dEtw + \tilde{\cal C} \,\right] \tbf \, ,
\end {equation}
where
\begin {equation}
    \dEtw =
    \cases {\displaystyle
	\frac{kT}{2\ppar\,(k{+}\ppar)_{\strut}} \left[ \tpp^2+\kappa \right]
	, & QCD; \cr \displaystyle
	\frac{kT}{2\ppar\,(k{+}\ppar)} \left[ \tpp^2+\kappa \right]
	+ {T \over 4k}
	\,, & QED, }
\label {eq:dEtw}
\end {equation}
and
\begin {equation}
    (\tilde {\cal C} \,\tbf)(\tpp)
    =
    \int \frac {d^2 \tqp}{(2\pi)^2} \;
    \tilde {\cal C}(\tqp)
    \Big[ \tbf(\tpp) - \tbf(\tpp{+}\tqp) \Big] \, ,
\end {equation}
with
\begin {equation}
    \tilde {\cal C}(\tqp)
    \equiv
    \kappa
    \int d\tilde \qpar \> d\tilde q^0 \;
    \delta (\tilde \qpar {-} \tilde q^0) \,
    \frac {1}{\tilde q}
    \left\{
	\frac{2}{\left|\tilde q^2 - \tilde\PiL(\tilde q^0,\tilde q)\right|^2}
	+
	\frac{[1-(\tilde q^0/\tilde q)^2]^2}
	    {\left|(\tilde q^0)^2-\tilde q^2
		-\tilde\PiT(\tilde q^0,\tilde q) \right|^2}
	\right\} .
\label {eq:C_dimless}
\end {equation}
Here $\tilde \qpar \equiv \qpar / \mD$, $\tilde q^0 \equiv q^0/\mD$,
and $\tilde q \equiv \sqrt{\tilde \q_\perp^2 + \tilde q_\parallel^2}$.
The explicit forms of the thermal gauge field self energies are
\cite{Weldon1}
\begin {eqnarray}
    \tilde\PiL(\tilde q^0,\tilde q)
    & = &
    -1 + \frac{\tilde q^0}{2\tilde q}
    \left[
	\ln \left(\frac{\tilde q{+}\tilde q^0}{\tilde q{-}\tilde q^0}\right)
	-i \pi
    \right] ,
\\
    \tilde\PiT(\tilde q^0,\tilde q)
    & = &
    \frac{(\tilde q^0)^2}{2\tilde q^2}
    +
    \left[\tilde q^2-(\tilde q^0)^2\right]
    \frac{\tilde q^0}{4 \tilde q^3}
    \left[
	\ln \left(\frac{\tilde q{+}\tilde q^0}{\tilde q{-}\tilde q^0}\right)
	- i\pi
    \right] .
\end {eqnarray}
These forms are valid for $\tilde q>|\tilde q^0|$,
which is enforced by the delta function in (\ref {eq:C_dimless}) which
identifies $\tilde q^0$ with $\tilde q_\parallel$.
Both $\dEtw$ and $\tilde {\cal C}$ should be viewed as linear operators
on the vector space of functions of transverse momentum.
Both operators are Hermitian --- in fact, real and symmetric --- 
under the inner product (\ref{eq:inner_prod}).

\subsection{Variational approach}

We are unable to find an exact solution of
\Eq{eq:f_dimless} for arbitrary values of $\kappa$
and $kT/\ppar(k{+}\ppar)$.
Instead we must use some approximative, but accurate, method.
We want a method which can be made
arbitrarily accurate as computational effort is increased, and which in
practice will have relative errors of less than $10^{-4}$.
This goal can be met with an adaptation of the
variational approach used in our previous work \cite{lead_log,largeN}.

First, separate the complex equation (\ref {eq:f_dimless}) into
a coupled pair of real equations,
\begin{equation}
\left[ \begin{array}{c} 2\tpp \\ 0 \\ \end{array} \right]
	= \left[ \begin{array}{cc} \tilde{\cal C} 
	& -\dEtw \\ -\dEtw & 
	-\tilde{\cal C} \\ \end{array} \right]
	\left[ \begin{array}{c} \Re\, \tbf \\
	\Im\, \tbf \\ \end{array} \right] \, .
\end{equation}
The linear operator represented by the $2 \times 2$ matrix
is real, symmetric, and non-singular.
(In fact, eigenvalues of this operator are bounded away from zero
by at least $|\kappa \, k T / [2\ppar (k{+}\ppar)]|$.)
Consider the functional 
\begin{equation}
	Q[\tbf] \equiv
	\Biggl(
	    \Big[ 2\tpp \,, 0 \Big] ,
	    \left[ \begin{array}{c} \Re\,\tbf \\ \Im\,\tbf \end{array} \right] 
	\Biggr)
	-
	{1\over2}
	\Biggl(
	    \Big[ \Re\, \tbf \,, \Im \, \tbf \Big] ,
	    \left[
		\begin{array}{cc} \tilde{\cal C} & -\dEtw \\
		-\dEtw & -\tilde{\cal C} \\ \end{array}
	    \right]
	    \left[ \begin{array}{c} \Re\,\tbf \\ \Im\,\tbf \end{array} \right] 
	\Biggr)
	\, .
\end{equation}
It possesses a unique extremum, at which \Eq{eq:f_dimless} is
satisfied.  Moreover, its value at the extremum is
\begin{equation}
Q_{\rm extremum} = \half \, \Big( 2\tpp \, , \, \Re\,\tbf \Big) \, ,
\end{equation}
which is precisely half the quantity
appearing in the final integral (\ref {eq:outer}) for the emission rate.
However, the extremum is a saddle-point,
rather than a maximum or minimum.
Nevertheless, if one uses a sufficiently flexible \ansatz\ for $\tbf$,
then the extremum under that \ansatz\ will be very close to the true extremum
and the determined value of $\Big( 2\tpp\, ,\, \Re\,\tbf \Big)$ should
be very accurate.
But because the extremum is a saddle-point,
a variational answer based on a finite basis set does not
automatically yield an upper or lower bound on the true answer,
nor is there any guarantee of monotonicity as the size of the
basis set is increased.

As noted earlier [{\em c.f.} \Eq {eq:rot_inv}], rotation invariance
implies that $\tbf(\tpp)$ equals $\tpp$ times some scalar function
$\chi(\tpsq)$.
So,
as an \ansatz\ for $\tbf$, we will choose a finite linear combination
of terms of this form,
\begin{equation}
    \Re\,\tbf(\tpp)
    = \tpp \sum_{m=1}^{N_{\rm r}} a_m \,\phi^{(m)}_{\rm r}(\tpsq) \, ,
\qquad
    \Im\, \tbf(\tpp)
    = \tpp \sum_{m=1}^{N_{\rm i}} b_m \,\phi^{(m)}_{\rm i}(\tpsq) \, .
\label{eq:trial}
\end{equation}
There is no need to choose the same basis of trial functions for the real and
imaginary parts.
Given the differing asymptotic behaviors
[{\em c.f.}, Eqs.~(\ref {eq:im_f1}) and (\ref {eq:re_f2})]
it is in fact sensible to use different basis sets for the real
and imaginary parts.
There is also
no need to require that the number of trial functions for the real and
imaginary parts, $N_{\rm r}$ and $N_{\rm i}$, be the same;
but in practice we will do so for simplicity.
Given this \ansatz, the functional $Q[\tbf]$ becomes a quadratic
function of the coefficients $\{ a_m \}$ and $\{ b_m \}$,
\begin{equation}
\label{eq:Q_ansatz}
Q[\tbf] = Q(\{a_m\},\{b_m\}) \equiv
	\sum_m a_m \, \tilde{S}_m
	- {1\over2} \sum_{m,n} \Big[ \, a_m \,, b_m \Big]
	\raisebox{-1.2ex}{$
	\left[ \begin{array}{cc} \tilde{C}^{\rm r}_{mn} & 
	-\dEtw_{mn}\vphantom{\Big|} \\ 
	-\dEtw_{nm}\vphantom{\Big|} & -\tilde{C}^{\rm i}_{mn}\\
	\end{array} \right]
	$}
	\raisebox{-1.2ex}{$
	\left[ \begin{array}{c} 
	a_n \vphantom{\Big(} \\
	b_n \vphantom{\Big(}
	\\ \end{array} \right]
	$}
	\, ,
\end{equation}
where
\begin{eqnarray}
\tilde S_m & \equiv & \Big( 2\tpp \, , \, \tpp \, \phi^{(m)}_{\rm r} \Big) \, ,
	\nonumber \\
\tilde C^{\rm r}_{mn} & \equiv & \Big( \tpp \, \phi^{(m)}_{\rm r} , \, 
	\tilde{\cal C} \: \tpp \, \phi^{(n)}_{\rm r} \Big) \, ,
	\nonumber \\
\tilde C^{\rm i}_{mn} & \equiv & \Big( \tpp \, \phi^{(m)}_{\rm i} , \, 
	\tilde{\cal C} \: \tpp \, \phi^{(n)}_{\rm i} \Big) \, ,
	\nonumber \\
\dEtw_{mn} & \equiv & \Big( \tpp \, \phi^{(m)}_{\rm r} , \,
	\dEtw \: \tpp \, \phi^{(n)}_{\rm i} \Big)  \,.
\end{eqnarray}
The extremal value of \Eq{eq:Q_ansatz} is given in terms of these
components by elementary linear algebra as,
\begin {equation}
    Q_{\rm extremum} = \half\,{\cal S}^{\rm T} \, {\cal M}^{-1} \, {\cal S} \,,
\end {equation}
where
\begin {equation}
    {\cal S} \equiv
    \left[ \begin{array}{c} \tilde{S}_n\vphantom{\Big(}\\
			0 \end{array} \right] , \qquad
    {\cal M} \equiv
	\left[ \begin{array}{cc}
	    \tilde{C}^{\rm r}_{mn} & -\dEtw_{mn} \vphantom{\Big|}\\ 
	    -\dEtw_{nm}\vphantom{\Big|} & -\tilde{C}^{\rm i}_{mn}
	\end{array} \right] ,
\end {equation}
are the indicated
$N_{\rm r}+N_{\rm i}$ component vector and (symmetric) matrix.

It remains to choose the trial functions and perform the integrals.
The choice of trial functions should be informed by examining
the general behavior of the integral equation (\ref{eq:f_dimless})
we are trying to solve.
At small $\tpsq$, $\chi(\tpsq)$ may be approximated as a
polynomial in $\tpsq$.  At large $\tpsq$, we have already determined its
behavior; from \Eq{eq:im_f1} and \Eq{eq:re_f2} one sees that the imaginary
part of $\chi(\tpsq)$ must go to zero as $1/\tpsq$ while the real part
falls as $1/(\tpsq)^3$.  We may choose trial functions so that
{\em every} trial function decreases at large $\tpsq$ at least as fast as
required; at least one trial function must fall with the correct
asymptotic power.
A nice set of trial functions which have the appropriate form is
\begin{eqnarray}
\phi^{(m)}_{\rm r}(\tpsq) &=& \frac{ 
	(\tpsq/A)^{m-1}}
	{\left( 1+\tpsq/A \right)^{N_{\rm r}+2}} \, ,
	\qquad m = 1, ..., N_{\rm r} \,,
\\
\phi^{(m)}_{\rm i}(\tpsq) &=& \frac{ 
	(\tpsq/A)^{m-1}}
	{\left( 1+\tpsq/A \right)^{N_{\rm i}}} \, ,
	\qquad \quad m = 1, ..., N_{\rm i} \,.
\end{eqnarray}
We have not specified the constant $A$, which should be chosen to be
close to the value of $\tpsq$ where $\tbf$ is varying in the most
significant way.  For instance, it can be chosen to be the value of
$\tpsq$ where $\Re \, \tbf$ reaches half of its asymptotic value, or the
value at which the integrand in $\Big( 2\tpp , \, \tbf\Big)$ is peaked.
The only difficulty is that the value of $\tpsq$ where $\tbf$ shows
significant behavior depends on $\kappa$, $k$, and $\ppar$.
We are interested in different values of $\kappa$ and
$k$, and $\ppar$ is integrated over in \Eq{eq:outer}.
Therefore a single value of $A$ will not give a good set of
trial functions for all values of $\kappa$, $k$, and $\ppar$.
Our solution is to determine all the matrix elements $\{ \tilde{C}_{mn} \}$ 
and $\{ \dEtw_{mn} \}$ for a variety of values of $A$.
Then we determine which value of $A$ to use, for a particular set
of parameters $[\kappa,k,\ppar]$,
based on a criterion which optimizes the sensitivity of the trial
functions to the most important range of $\tpsq$.  
Various specific criteria could be used; we chose the following.
We want the integral
\begin{equation}
\Big( 2\tpp \, , \, \tbf \Big) = \frac{1}{2\pi} \int_0^\infty
	\Re \, \chi(\tpsq) \, \tpsq \, d(\tpsq)
\end{equation}
to be accurately determined.
Map the integral to a finite interval by defining
\begin{equation}
    y \equiv \frac{|\tpp|}{A^{1/2}+|\tpp|} \, ,
\end{equation}
and rewrite the integral in terms of $y$,
\begin{equation}
\Big( 2 \tpp \, , \, \tbf \Big) = 
    \frac{1}{\pi} \int_0^1
	\frac{ A^2 \, y^3} {(1{-}y)^5} \;
	\chi \Bigl( Ay^2/(1{-}y)^2 \Bigr) \; dy \, .
\label {eq:y_int}
\end{equation}
For each value of $A$, 
we solve for the variational coefficients $ \{ a_m,b_m \}$ and determine
the resulting function $\chi(\tpsq)$.
We then choose the value of $A$ for which the maximum of the integrand
of \Eq{eq:y_int} occurs closest to $y=1/2$, which is where our
basis functions have the most sensitivity.%
\footnote
    {%
     The maximizing value of $y$ sometimes depends on $A$ such that it
     crosses $y=1/2$ not only at $A \sim \sqrt{1+(k[k{+}p]/kT)}$,
     but also at a very large value of $A$.  This second solution
     is spurious and should be discarded.
    }
This ensures that we are always using a set of trial functions
which are suitably flexible in the region most important
for evaluating the integral that we care about.

The final integral over $\ppar$ in \Eq{eq:outer} is one dimensional,
and performing it by numerical quadrature is elementary.
Doing so completes the calculation of the contribution of bremsstrahlung
and \ipa\ to the photo-emission rate.

\subsection{Integrals for $\tilde{C}_{mn}$}

The integrals determining $\dEtw_{mn}$ and $\tilde{S}_m$
are elementary for our choice of trial functions and can be done in
closed form.
One finds
\begin {eqnarray}
    \tilde S_m &=& 2 A^2 \, K(m,N_{\rm r}) \,,
\\
    \dEtw_{mn}
    &=&
    {k A^2 T \over 2\ppar (k{+}\ppar)}
    \left[
	A \, K(m{+}n,N_{\rm r}{+}N_{\rm i})
	+
	\eta \, K(m{+}n{-}1,N_{\rm r}{+}N_{\rm i})
    \right] ,
\end {eqnarray}
where
$K(j,N) \equiv {j! \, (N{-}j)! / [4\pi \, (N{+}1)!]}$
and
$\eta = \kappa$ for QCD, or $\kappa + \half \, \ppar (k{+}\ppar) / k^2 $
for QED.
%
The collision integrals $\{ \tilde{C}_{mn} \}$ cannot be evaluated
analytically, but may be reduced to a numerically tractable form as follows.
The integral we need is
\begin{eqnarray}
\tilde{C}_{mn} & = &
	\half \!\int\! \frac{d^2 \tpp}{(2\pi)^2}
	\int \! \frac{d^2 \tqp}{(2\pi)^2}  \;
	\tilde{\cal C}(\tqp)
\times {}
\nonumber\\ && 
	\left[
	    \tpp \phi^{(m)}(\tpsq) - (\tpp{+}\tqp) \phi^{(m)}(|\tpp{+}\tqp|^2)
	\right] 
	\cdot
	\left[
	    \tpp \phi^{(n)}(\tpsq) - (\tpp{+}\tqp) \phi^{(n)}(|\tpp{+}\tqp|^2)
	\right] .
\nonumber\\
\end{eqnarray}
Here we have used the symmetry of the collision operator to
write the integrand in a form manifestly symmetric under $m \leftrightarrow n$,
which also makes apparent that the collision operator is positive definite
on all normalizable functions $\tbf$.
One overall angular integral
may be performed, and the remaining integrals become
\begin{eqnarray}
\tilde{C}_{mn} & = & \frac{1}{32 \pi^2} \int_0^\infty 
	d(\tilde{q}_\perp^2) \; \tilde{\cal C}(\tilde q_\perp) \,
	I_p(\tilde{q}_\perp) \, ,
\\
I_p(\tilde{q}_\perp) & = &
	\int_0^\infty \! d(\tilde{p}_\perp^2)
	\int_{-\pi}^{\pi} \frac{d\theta}{2\pi}
	\bigg\{\,
	    \tpsq \, \phi^{(m)}(\tpsq) \, \phi^{(n)}(\tpsq)
	    + |\tpp{+}\tqp|^2
	    \phi^{(m)}(|\tpp{+}\tqp|^2) \, \phi^{(n)}(|\tpp{+}\tqp|^2)
\nonumber \\ && \hspace{1.05in} {}
	    - \Big(\tpsq{+}\tilde{p}_\perp \tilde{q}_\perp \cos\theta \Big)
	    \Big[
		\phi^{(m)}(|\tpp{+}\tqp|^2) \, \phi^{(n)}(\tpsq) +
		(m \leftrightarrow n)
	    \Big]
	\bigg\} \,,
\end{eqnarray}
where in the last equation $(\tpp{+}\tqp)^2$ is shorthand for
$\tpsq + \tqsq + 2\tilde{p}_\perp \tilde{q}_\perp \cos \theta$.
Although we are not able to evaluate either
$I_p(\tilde q_\perp)$ or $\tilde{\cal C}(\tilde q_\perp)$ analytically,
since no more than three integrals ever nest it is not difficult
to perform all the integrals by numerical quadrature,
using adaptive mesh refinement, and achieve
better than a part per million relative accuracy.%
\footnote
    {%
    When trying to achieve high accuracy, there can be
    some difficulty in the $\theta$ integral near $\theta=\pi$
    when $\tilde{p}_\perp$ is
    very close to $\tilde{q}_\perp$ and both are large.  This can be
    eliminated by noting that the integral has a factor of two redundancy;
    the integral we want is twice the integral where $\theta$ is restricted
    to values for which $(\tpp{+}\tqp)^2 > \tpsq$.
    }

We have evaluated the quantities
$\{ \tilde C^{\rm r}_{mn} \}$ and $\{ \tilde C^{\rm i}_{mn} \}$,
for several values of $A$ in the range $0.6$ to $20$,
using $N_{\rm i} = N_{\rm r} \equiv N$ ranging from 2 up to 16.
Reducing $N$ from 10--16 to 6--10 
changes our results for the integral (\ref{eq:outer}) 
by less than 50 parts per million for all values of
$k$ and $\kappa$ which we consider ($k\geq T/10$).  
Using only 4--6 trial functions causes errors of less than $0.2\%$,
while using 2--4 trial functions allows errors of a few percent.
We conclude that this variational approach, with a modest set of
test functions and flexible selection of the overall scale $A$, yields
a very accurate determination of $\Big( \tpp , \tbf \Big)$,
and likewise for the final bremsstrahlung and \ipa\ rates.

\section{$2\leftrightarrow 2$ processes beyond leading-log}
\label{sec:K}

As discussed in the Introduction,
bremsstrahlung and \ipa\ processes are not the only
processes which lead to photo-emission at order $\alphas \alphaEM$.
There are also the $2\leftrightarrow 2$ particle processes depicted in
Fig.~\ref{fig:2to2}.  Neglecting for a moment the subtleties
associated with the kinematic region where the internal propagator
becomes soft,
and ignoring the (small)
corrections to the external state dispersion relations due to
both thermal effects and explicit mass terms,
the contributions to the photon production rate from
(gluon-photon) Compton scattering and quark-antiquark annihilation are
\begin{eqnarray}
\label{eq:compton1}
\Gamma_\gamma^{\rm Compton} & = & 
	\int \frac{d^3 \p \> d^3 \p' \> d^3 \k \> d^3 \k'}
	{(2\pi)^{12} \, 16 \, p \, k \, p' \, k'} \>
	(2\pi)^4 \, \delta^4(p_\mu{+}p'_{\mu}{-}k_\mu{-}k'_\mu) \>
	16 e^2({\textstyle \sum_{s} q_{s}^2})
\times \nonumber \\ && \hspace{1in} {} \times
	\df \, \cf \, \gs^2 \left[ \frac{-s}{t} + \frac{-t}{s} \right]
	n_f(p) \, n_b(p') \, [1{-}n_f(k')] \, ,
\\
    \Gamma_\gamma^{\rm Annihilation} & = & 
	\int \frac{d^3 \p \> d^3 \p' \> d^3 \k \> d^3 \k'}
	{(2\pi)^{12} \, 16 \, p \, k \, p' \, k'} \>
	(2\pi)^4 \, \delta^4(p_\mu{+}p'_{\mu}{-}k_\mu{-}k'_\mu) \>
	16 e^2 ({\textstyle \sum_{s} q_{s}^2})
\times \nonumber \\ && \hspace{1in} {} \times
	\df \, \cf \, \gs^2 \left[ \frac{u}{t} \right]
	n_f(p) \, n_f(p') \, [1{+}n_b(k')] \, , 
\label{eq:annih1}
\end{eqnarray}
respectively,
with $\p$ and $\p'$ the incoming momenta, $\k$ the photon momentum,
$\k'$ the other outgoing momentum, 
$s$, $t$, and $u$ the usual Mandelstam variables,
and $p\equiv |\p| = p^0$, etc.

These expressions must be corrected when the momentum on an internal
propagator becomes small, because thermal effects then significantly
modify the matrix element.
This occurs for the $u/t$ and $s/t$ terms when $t$ is of order $g^2 T^2$.
Therefore it is best to arrange these integrations so that
$|\p{-}\k|$ appears as an integration variable,
and to separate off the soft (small $|\p{-}\k|$) region for
special treatment.  Such a separation and treatment of the soft region
was done correctly in Refs.~\cite{Kapusta,Baier}.
Outside this soft region,
\Eq{eq:compton1} and \Eq{eq:annih1} are correct at leading order in $\gs$.
The large $|\p{-}\k|$ region was treated in Refs.~\cite{Kapusta,Baier}
only in the limit $k \gg T$, but we will 
not make any such restriction in our evaluation here.

Because the population functions take a simple form in the plasma rest frame,
we choose to work in terms of plasma frame quantities.
We first consider the term in \Eq{eq:compton1} involving the
$t/s$ matrix element.
This term requires no special handling of the soft region,
and is most easily done in the ``$s$-channel'' parameterization of
Ref.~\cite{largeN}.%
\footnote{
	Our labeling of external momenta differs from our previous
	work \pcite{lead_log,largeN}.
	The incoming momenta are now
	labeled $(\p,\p')$ and the outgoing momenta are $(\k,\k')$,
	whereas in the papers \pcite{lead_log,largeN}
	the incoming momenta were $(\p,\k)$ and 
	outgoing momenta were $(\p',\k')$.  The current choice assigns
	the photon momentum $k$, consistent with 
	Ref.~\pcite{us} and the previous section.
	}
Defining $\omega\equiv p{+}p'$ and $q\equiv |\p{+}\p'|$,
the photo-production rate resulting from the $t/s$ term is
\begin{eqnarray}
    \Gamma_\gamma^{[t/s]} & = &
	\frac {\df \, \cf \, \gs^2 e^2{\textstyle (\sum_{s} q_{s}^2)}}
	{(2 \pi)^6}
	\int_0^\infty \! d\omega \int_0^\omega {dq \over q^2}
	\int_{\frac{\omega{-}q}{2}}^{\frac{\omega{+}q}{2}} dp
	\int_{\frac{\omega{-}q}{2}}^{\frac{\omega{+}q}{2}} dk
	\int_0^{2\pi} \! d\phi \; 
\times \\ && \hspace{0.8in} {} \times
	n_f(p) \, n_b(\omega{-}p) \, [1{-}n_f(\omega{-}k)] \;
\times \nonumber \\ && \hspace{0.8in} {} \times
	\left\{ q^2-(2p{-}\omega)
	(2k{-}\omega) - \cos \phi
	\sqrt{[q^2{-}(\omega{-}2p)^2] [q^2{-}(\omega{-}2k)^2]} \, \right\} .
\nonumber
\end{eqnarray}
The $\phi$ integration is trivial and just eliminates the $\cos\phi$ term.
To determine the energy dependence
of the photon spectrum we want the result at a fixed value
of $k$, which is
\begin{equation}
    \frac{d\Gamma_\gamma^{[t/s]}}{dk} =
    {k^2 {\cal A}(k) \over 2\pi^2} \; I_1(k) \,,
\end{equation}
with
\begin{equation}
	I_1(k)
	\equiv
	\frac {1} {2\pi^2 T^2} \!
	\int_k^\infty \! d\omega
	\int_{|2k-\omega|}^\omega {dq \over q^2}
	\int_{\frac{\omega{-}q}{2}}^{\frac{\omega{+}q}{2}} dp \;
	\frac{n_f(p) \, n_b(\omega{-}p) \, [1{-}n_f(\omega{-}k)]} {k \, n_f(k)}
	\left[ q^2 - {(2p{-}\omega)(2k{-}\omega)} \right] .
\label{eq:K_from_s}
\end{equation}
We have factored out the same leading logarithmic coefficient
${\cal A}(k)$ defined in \Eq{eq:calA}, in such a way that
$I_1(k)$ directly gives a contribution to $\Ctwo(k/T)$.

Next we turn to the terms involving $s/t$ and $u/t$ matrix elements.
For these terms, let $\omega \equiv (k{-}p)$ and $q \equiv |\k{-}\p|$.
Introduce a separation scale $q_*$ satisfying $\gs T \ll q_* \ll T$
and, as mentioned above,
separate the soft ($q < q_*$) contributions from the remainder.
The emission rate from the soft $q < q_*$ part of the $s/t$ and $u/t$ terms
requires the inclusion of thermal corrections on internal propagators,
which modifies the matrix element from what is written in
Eqs.~(\ref {eq:compton1}) and (\ref {eq:annih1}).
However, one can make $q\ll T$ approximations which
render the integrals tractable.  The $q<q_*$ result  
was found in Ref.~\cite{Baier}
and is%
\footnote{Eq.~(17) of Ref.~\pcite{Baier} leaves one un-evaluated integral,
	which when performed gives this result.
        Ref.~\pcite{Kapusta}
	computes the $(q^2 - \omega^2)<q_*^2$ contribution, which can
	also be converted to the result quoted.}
\begin{equation}
\label{eq:smallq}
\left. \frac{d\Gamma_\gamma^{[u/t , s/t]}}{dk} \right|_{q < q_*} =
	\frac{k^2 {\cal A}(k)}{2\pi^2}
	\left[ \,
	    \ln \left(\frac{q_*}{m_\infty}\right) - 1 + \ln 2 \,
	\right] .
\end{equation}
The emission rate from the remaining $q > q_*$ part of the $s/t$ and $u/t$
terms is
\begin{eqnarray}
\label{eq:su_const}
\left. \Gamma_\gamma^{[u/t , s/t]} \right|_{q > q_*} & = &
	\frac{\df \, \cf \, \gs^2 e^2{\textstyle (\sum_{s} q_{s}^2)}}
	{(2\pi)^6}
	\int_{q_*}^\infty {dq \over q^2}
	\int_{-q}^{q} \! d\omega
	\int_{\frac{q{-}\omega}{2}}^{\infty} dp
	\int_{\frac{q{+}\omega}{2}}^{\infty} dp'
	\int_0^{2\pi} \! d\phi \;
\times \nonumber \\ && {} \times
	\biggl\{ n_f(p) \, n_b(p') \,[1{-}n_f(p'{-}\omega)]
	\left[ (2p{+}\omega)(2p'{-}\omega)+q^2 - \# \cos \phi \right]
+ \nonumber \\ && \!\quad {} +
	n_f(p) \, n_f(p') \, [1{+}n_b(p'{-}\omega)]
	\left[ (2p{+}\omega)(2p'{-}\omega)-q^2 -\# \cos \phi \right]
	\biggr\} \, ,
\end{eqnarray}
where we have not written the coefficient on the $\cos \phi$ pieces
because the $d\phi$ integral simply removes these terms.
We wish to extract the part of this rate corresponding to
fixed $k = p{+}\omega$.
The result is
\begin{equation}
    \left.  \frac{d\Gamma_\gamma^{[u/t,s/t]}}{dk} \right|_{q > q_*}
    =
    {k^2 {\cal A}(k) \over 2\pi^2} \; I_2(k,q_*) \,,
\end {equation}
with
\begin{eqnarray}
	I_2(k,q_*)
    &\equiv&
	\frac{1} {2\pi^2 T^2}
	\int_{q_*}^\infty \frac{dq}{q^2}
	\int_{-q}^{{\rm min}(q,2k{-}q)} d\omega
	\int_{\frac{q{+}\omega}{2}}^{\infty} dp'
\nonumber
\times \\ && \quad\qquad {} \times
	\left\{
	    \frac{n_f(k{-}\omega) \, n_b(p') \, [1{-}n_f(p'{-}\omega)]}
		    {k\,n_f(k)}
	    \left[ (2k{-}\omega)(2p'{-}\omega) + q^2 \right]
+  \right. \nonumber \\ && \left. \quad\qquad \;\;\; {} +
	    \frac{n_f(k{-}\omega) \, n_f(p') \, [1{+}n_b(p'{-}\omega)]}
		    {k\,n_f(k)}
	    \left[ (2k{-}\omega)(2p'{-}\omega) - q^2 \right]
	\right\} .
\label{eq:K_really}
\end{eqnarray}
The complete contribution to the emission rate
from these $2 \leftrightarrow 2$ processes
has the form given in \Eq{eq:2to2} with
\begin {equation}
    \Ctwo(k/T) = \lim_{q_*\to0}
    \left[
	I_1(k) + I_2(k,q_*) + \ln (q_*/T) + \half \ln (2T/k) -1
    \right] .
\label {eq:Ctwo}
\end {equation}
For $q \ll T$, the $\omega$ and $p'$ integrals in \Eq{eq:K_really}
may be performed analytically and yield
$
    I_2(k,q_*) \sim \int_{q_*} dq/q = -\ln q_* + {\rm const.}
$,
which confirms that the $q_* \to 0$ limit in the
full result (\ref {eq:Ctwo}) is well behaved.
If one assumes that $k \gg T$, then the remaining integrals in
Eqs.~(\ref{eq:K_from_s}) and (\ref{eq:K_really})
can also be evaluated in closed form,
to leading order in $T/k$,
yielding the asymptotic result for $\Ctwo$ presented earlier
in \Eq{eq:K_is}.
For general $k \sim T$,
one must perform the triple integrals in Eqs.~(\ref{eq:K_from_s}) and
(\ref{eq:K_really}) by numerical quadrature.
To obtain accurate results,
one should use several values of $q_*$ and perform an
extrapolation to small $q_*$ to remove $O(q_*/T)$ corrections.
This we have done.
The integrals are not numerically demanding. 

\newpage
\section{Results}
\label{sec:results}

\begin{figure}[htp]
\centerline{\epsfxsize=3.5in\epsfbox{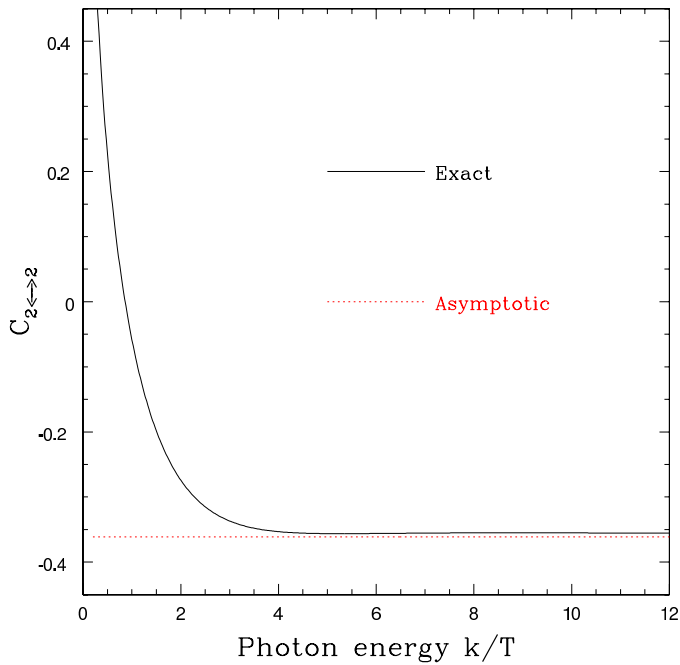}}
\vspace{0.1in}
\caption{ \label{fig:C_2to2}
    Contribution $\Ctwo(k/T)$ to the constant under the log
    arising from $2\leftrightarrow 2$ particle processes
    (gluon-photon Compton scattering and quark-antiquark annihilation),
    as a function of $k/T$.
    The dotted line is the asymptotic limit, derived by
    \protect\cite{Kapusta,Baier}.
    The result is quite close to the asymptotic value already at $k/T = 3$;
    the visible but tiny difference between the curve and
    the asymptotic value is due to a $1/k$ tail with a very small coefficient.
    }
\end{figure}

The complete leading-order spontaneous photon emission rate may be written as
\begin{equation}
    \GammaE(\k) \equiv
    (2\pi)^3 \, {d\Gamma_\gamma \over d^3\k} =
    {\cal A}(k) \,
    \Bigl[ \, \ln \left({T}/{m_\infty}\right) + C_{\rm tot}(k/T) \, \Bigr] \,,
\label {eq:final_rate}
\end {equation}
with the coefficient ${\cal A}(k)$ given in \Eq{eq:calA} and
\begin {equation}
    C_{\rm tot}(k/T) \equiv \half \ln \left({2k}/{T}\right)
    + \Ctwo(k/T) + C_{\rm brem}(k/T) + C_{\rm annih}(k/T)
\label {eq:Ctot}
\end{equation}
the total constant under the log.
The contribution $\Ctwo(k/T)$ is determined from the sum of
Eqs.~(\ref{eq:K_from_s}) and (\ref{eq:K_really}).
It is independent of $\kappa
\equiv m_\infty^2/\mD^2$, and is presented in Fig.~\ref{fig:C_2to2},
which also shows the asymptotic value (\ref {eq:K_is})
derived in Refs.~\cite{Kapusta,Baier}.
Note that the asymptotic result provides quite a good approximation provided
$k \gsim 3\,T$.

\begin{figure}[tp]
\centerline{\epsfxsize=3.2in\epsfbox{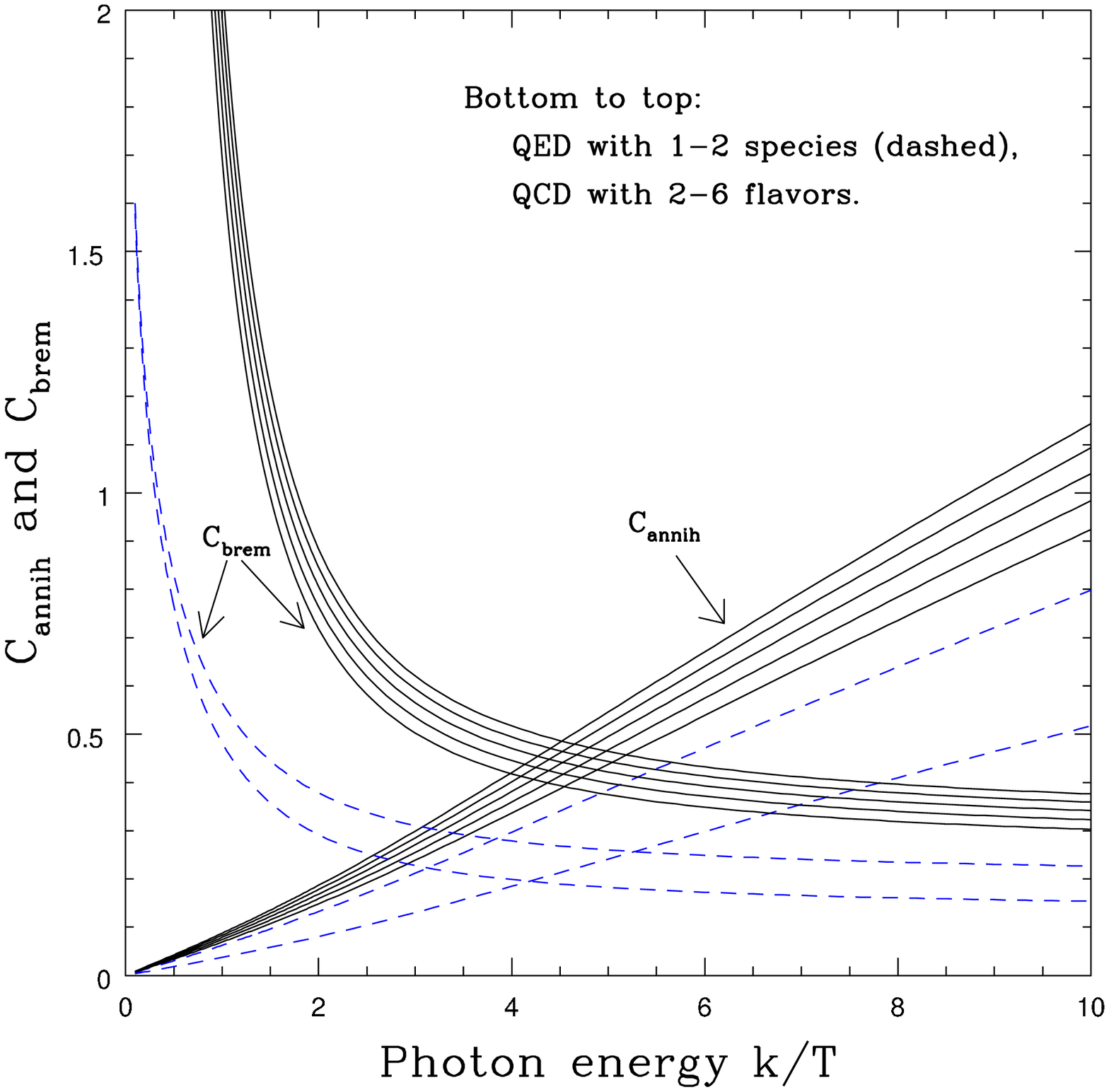} \hspace{0.1in}
\epsfxsize=3.2in\epsfbox{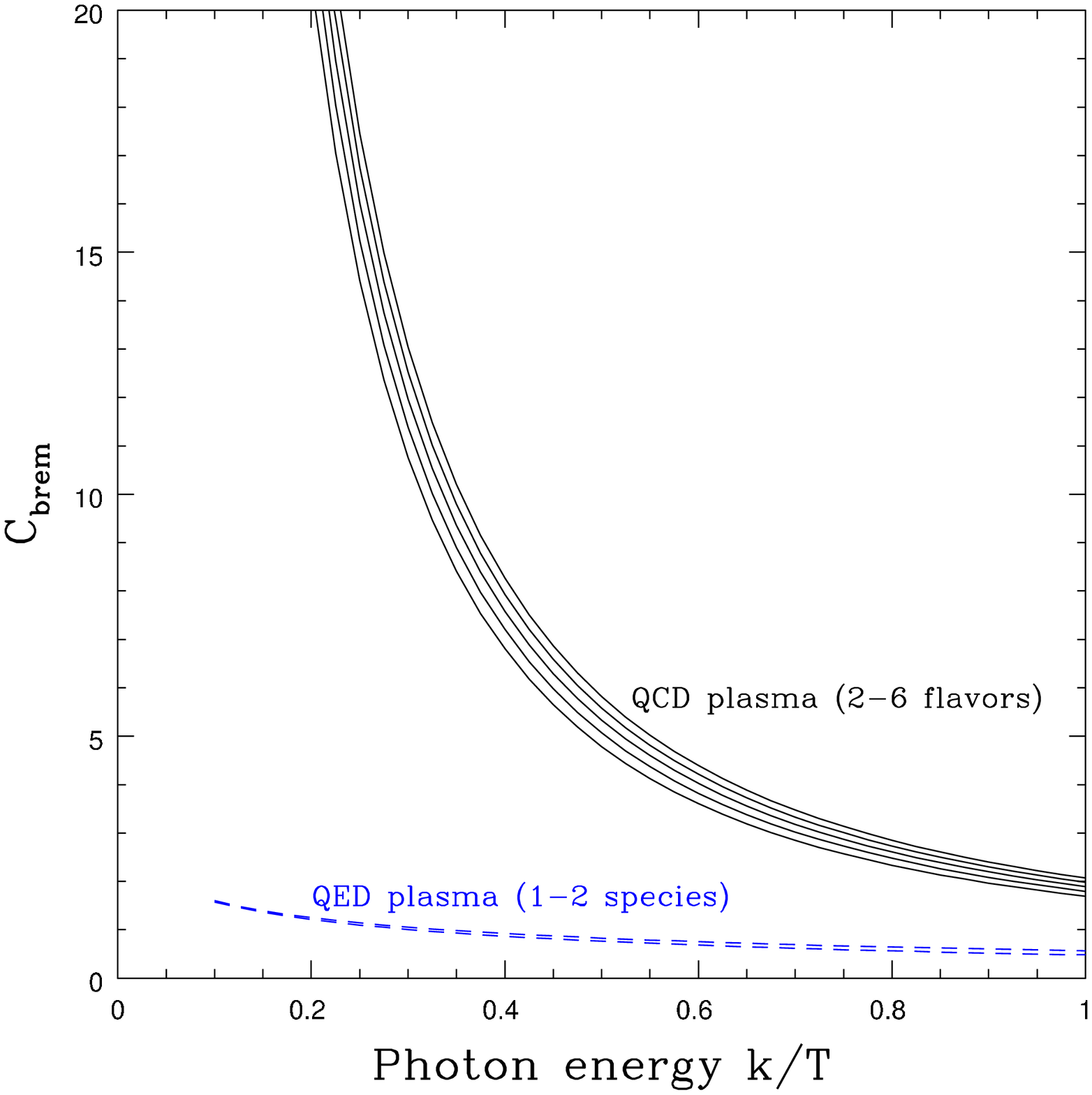}}
\vspace{0.1in}
\caption{\label{fig:C_brem}
    Left panel:
    Contributions $C_{\rm brem}(k/T)$ and $C_{\rm annih}(k/T)$ to the
    constant under the log arising from near-collinear
    bremsstrahlung and pair production.
    The curves rising at small $k/T$ are plots of $C_{\rm brem}(k/T)$,
    while the curves rising at large $k/T$ are plots of
    $C_{\rm annih}(k/T)$.
    In both cases, the curves from bottom to top
    are for $\kappa = {}$3/4, 3/8, 1/4, 2/9, 1/5, 2/11, and 1/6,
    corresponding to electron plasma, electron plus muon plasma,
    and QCD with 2, 3, 4, 5, or 6 quark flavors, respectively.
    In the bottom two curves, we include the plasma induced photon
    dispersion, relevant for QED plasmas, which suppresses the
    soft bremsstrahlung rate.
    Right panel:
    Magnified view of the soft photon region of
    the bremsstrahlung contribution.
    }
\end{figure}

The functions $C_{\rm brem}(k/T)$ and $C_{\rm annih}(k/T)$
are $\kappa$ dependent.
We show $C_{\rm brem}(k/T)$ [defined as the contribution to the integral
(\ref{eq:outer_int}) from the regions $\ppar>0$ plus $\ppar<-k$],
and $C_{\rm annih}(k/T)$
[the $-k<\ppar<0$ contribution to \Eq{eq:outer_int}]
separately in Fig.~\ref{fig:C_brem},
for a number of physically interesting values of $\kappa$.
The values shown were selected to equal the
value for an Abelian electron plasma ($\kappa=3/4$), an electron plus
muon plasma ($\kappa=3/8$), and a quark-gluon plasma with 2 to 6
flavors ($\kappa=2/(6+\nf)$).  
For the Abelian plasmas, the photon dispersion correction is included
in $\delta E$ [{\em c.f.} \Eq{eq:dEtw}],
which suppresses soft bremsstrahlung.

\begin{figure}[htp]
\centerline{\epsfxsize=3.5in\epsfbox{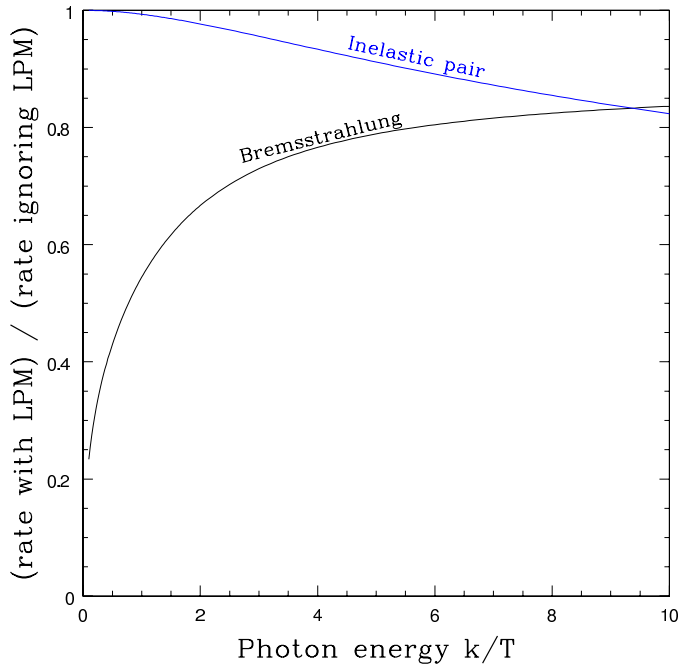}}
\vspace{0.1in}
\caption{ \label{fig:lpm}
    Relative importance of the LPM effect.
    The figure shows the ratio of the true bremsstrahlung and \ipa\ 
    rates to the rates which result if one neglects the LPM effect
    in the treatment, for the case of $\kappa = 1/4$
    which corresponds to two flavor QCD.
    The LPM suppression is rather modest, 30\% or less,
    except for bremsstrahlung photons of energy $k \lsim 2\,T$.
    These plots show very little change
    for other values of the mass ratio $\kappa$.
}
\end{figure}

The effect of LPM suppression on the bremsstrahlung and \ipa\ rates
is shown in Fig.~\ref{fig:lpm}, which plots
the ratio of the actual bremsstrahlung and \ipa\ rates to
the corresponding rates neglecting the LPM effect,
for the case of $\nf = 2$ QCD.
For most momenta, it is evident that the LPM suppression is
a rather modest 35\% or less.
The exception, for the range of photon momenta shown ($k<10\,T$),
is bremsstrahlung with $k\lsim 2\,T$.  Here the LPM effect is important.
It also makes a large relative suppression to the rate
for \ipa\ at very large $k/T$,
but in practice this means $k>20 \,T$,
where the statistical factor $n_f(k) \sim e^{-k/T}$
ensures that virtually no photons are produced.

\begin{figure}[htp]
\centerline{
\epsfxsize=3.2in\epsfbox{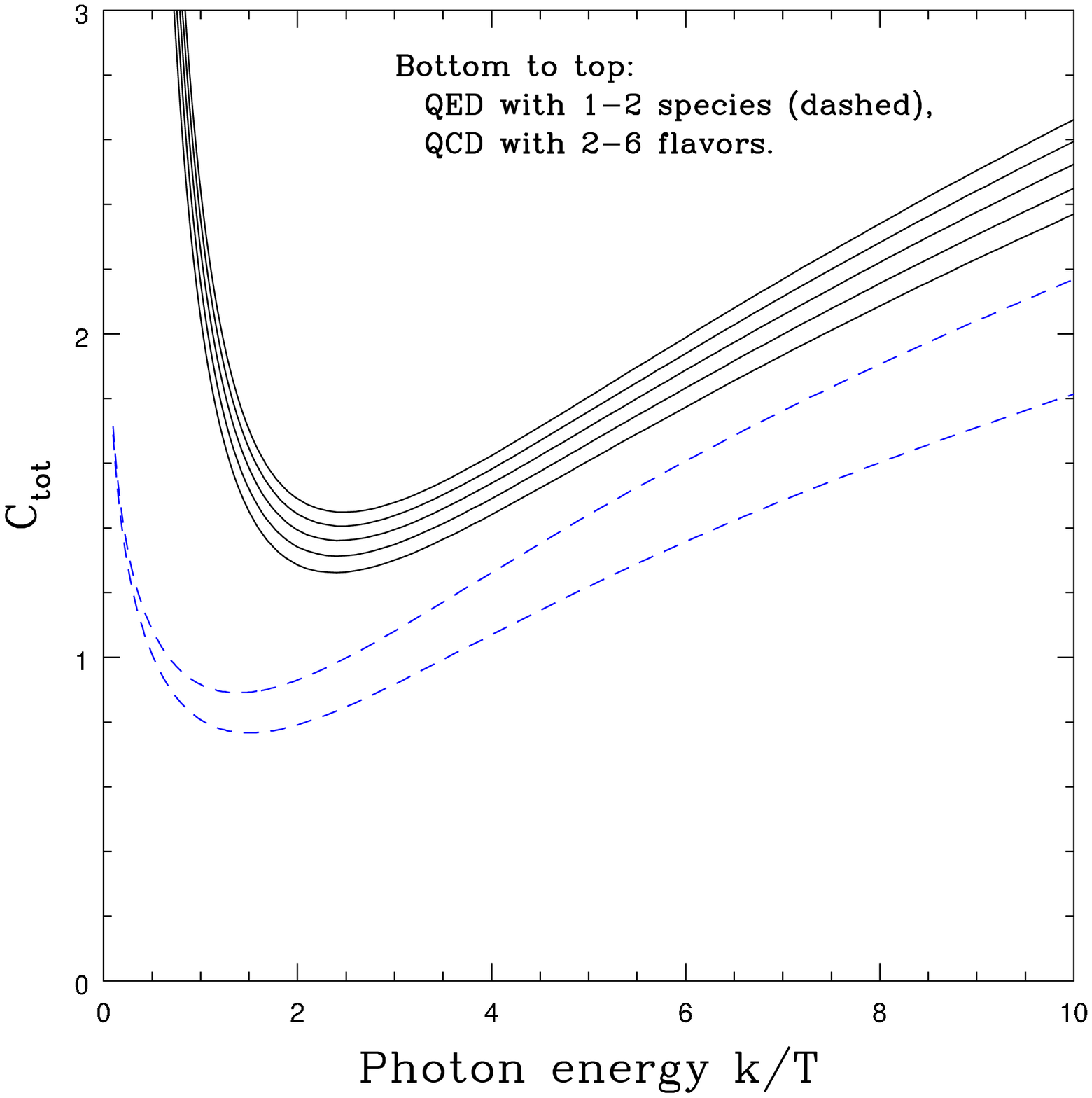} \hspace{0.1in}
\epsfxsize=3.2in\epsfbox{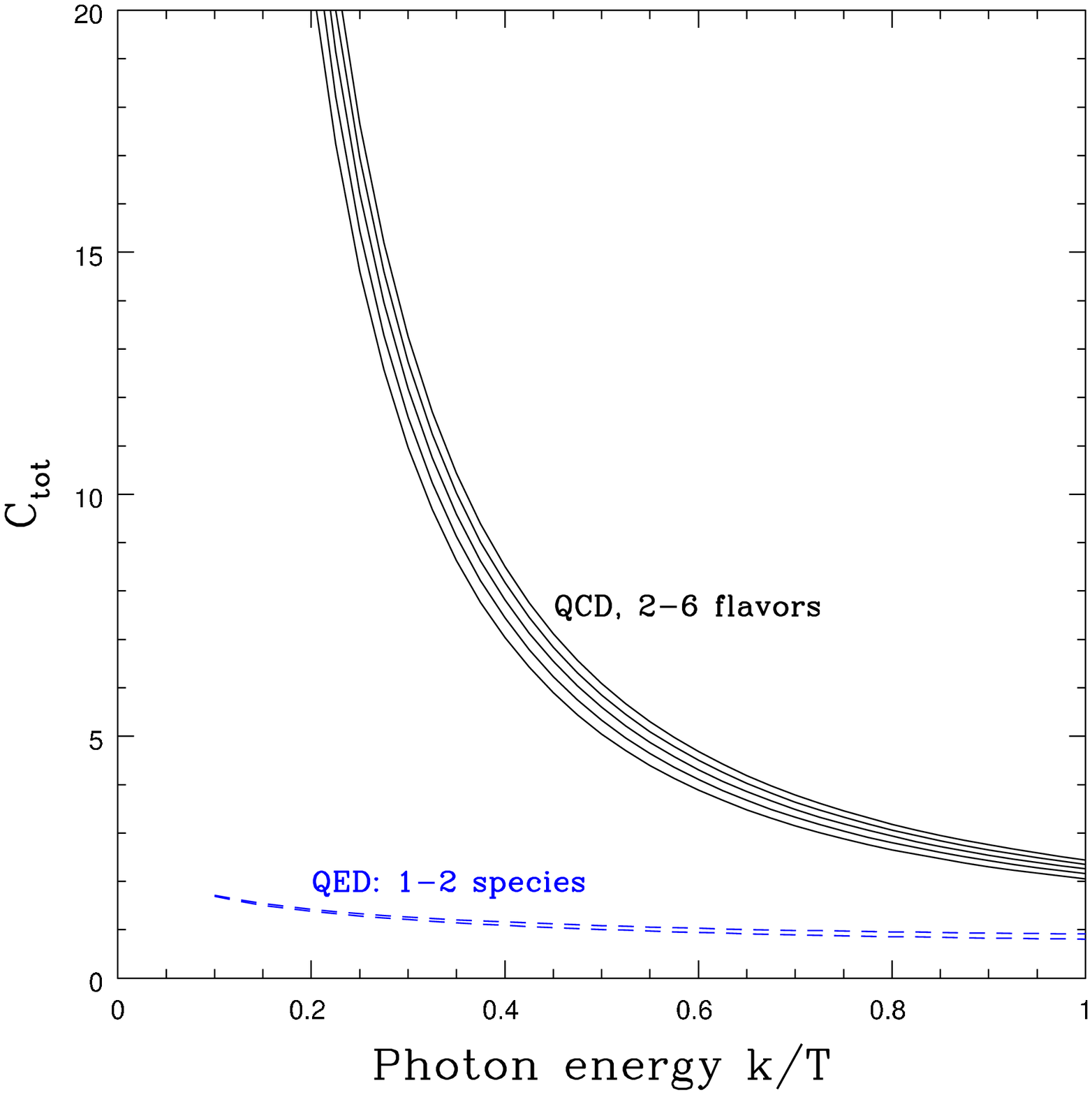}}
\vspace{0.1in}
\caption{\label{fig:C_summed}
    Left panel:
    Total constant under the log, equal to the sum of contributions
    $
    C_{\rm tot}(k/T) \equiv$ $ 
    \half \log(2k/T) + \Ctwo(k/T) + C_{\rm brem}(k/T) + C_{\rm annih}(k/T)
    $.
    This is the value which must be added to $\ln (T/m_\infty)$
    to obtain the total emission rate,
    as indicated in Eq.~(\protect\ref{eq:final_rate}).
    The different curves (from bottom to top)
    are the same as in Fig.~\protect\ref{fig:C_brem}.
    Right panel:
    Magnified view of the soft photon region.}
\end{figure}

The complete constant under the log,
$C_{\rm tot}(k/T)$,
equal to the sum of contributions shown in \Eq{eq:Ctot},
is plotted in Fig.~\ref{fig:C_summed}.
Over a wide range of photon energies, $T \lsim k \lsim 6T$,
one sees that $C_{\rm tot}(k/T)$ lies between about 1 and 2.

\begin{figure}[tp]
\centerline{
\epsfxsize=3.2in\epsfbox{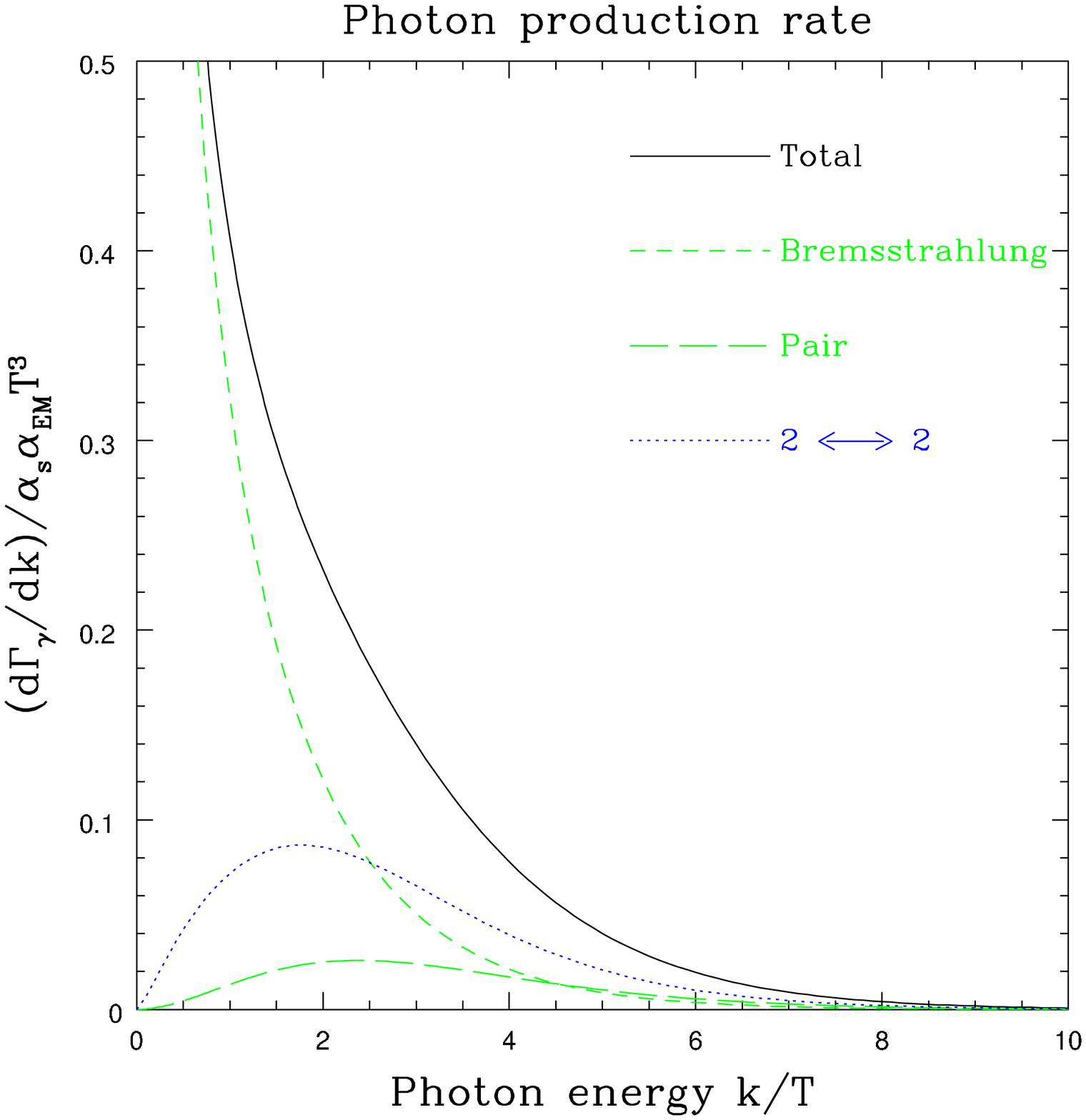} \hspace{0.1in}
\epsfxsize=3.2in\epsfbox{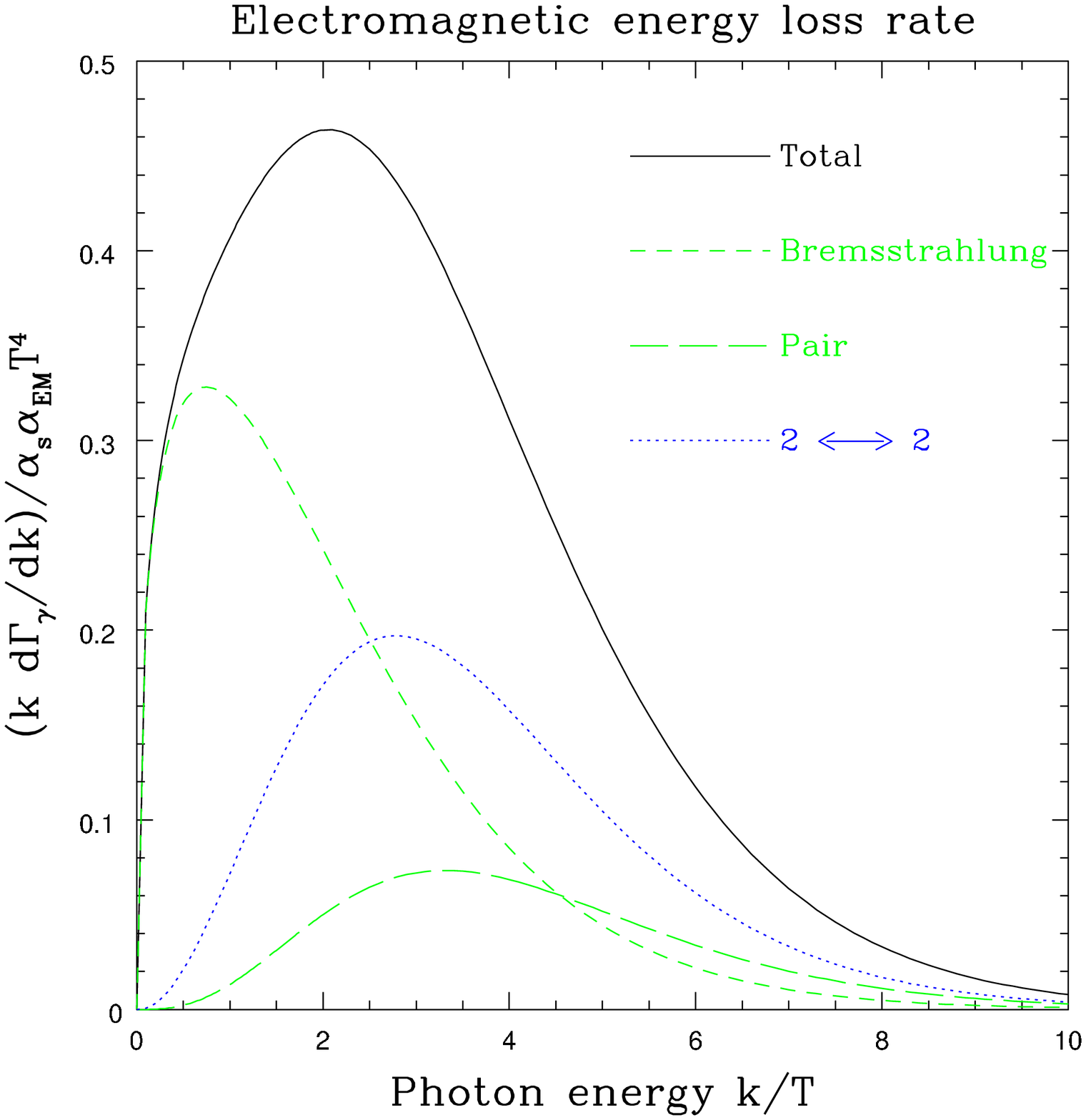}}
\vspace{0.1in}
\caption{\label{fig:total_rates}
    Total photon emission rate,
    together with the bremsstrahlung, \ipa\,
    and $2\leftrightarrow 2$ contributions,
    for two-flavor QCD with $\alphas=0.2$.
    The left panel shows $d\Gamma_\gamma/dk$,
    divided by $\alphas \, \alphaEM \, T^3$,
    while the right panel shows rates weighted by photon energy.}
\end{figure}

Finally, Fig.~\ref{fig:total_rates} shows the full leading-order
photon emission rate,
together with the bremsstrahlung, \ipa\, and $2\leftrightarrow 2$
contributions, for two-flavor QCD with $\alphas=0.2$.
The left panel shows the production rate $d\Gamma_\gamma /dk$,
while the right panel shows the rate weighted by a factor of the
photon energy.
One sees that bremsstrahlung is the
dominant production mechanism for not-so-hard photons with $k < 2T$.
As discussed in section \ref {sec:lpm},
the ratio of the bremsstrahlung rate to
the $2 \leftrightarrow 2$ rate scales, at
small $k$, as $(T/k)^{3/2}$ (up to logs),
a factor of $\sqrt{k/T}$ less strongly than
if there were no LPM effect.
At intermediate values of $k/T \gsim 4$, where
the most experimentally detectable photons are probably produced, 
the $2 \leftrightarrow 2$ processes generate the largest contribution,
comparable to the sum of the two near-collinear inelastic processes.
At large $k \gg T$, the \ipa\ emission rate grows,
in comparison to the $2 \leftrightarrow 2$ rate,
almost linearly with $k/T$ up to quite large $k$.
As shown in section \ref {sec:lpm},
the ratio of these rates ultimately grows only as $(k/T)^{1/2}$
but this asymptotic behavior turns out to set in only for $k>10 \,T$.

\section{Conclusions}

The above results represent
the first complete leading order calculation of the
photon emission rate from a hot, weakly coupled, thermalized quark gluon
plasma at zero chemical potential.  In addition to well known
$2 \leftrightarrow 2$ particle processes,
near-collinear bremsstrahlung and \ipa\ also make
leading order contributions.
A consistent treatment of these latter processes requires a
rather complicated analysis in order to incorporate correctly
the effects of multiple soft scatterings which may occur
during the emission of the photon.
These lead to an $O(1)$ suppression of these near-collinear processes.
To account for this suppression, known as the LPM effect,
we solved the integral equation, derived in Ref.~\cite{us},
which accounts for partial coherence between successive scattering events.

For modest values of $\alphas \sim 0.2$,
we find that near-collinear bremsstrahlung
dominates the photo-emission rate for soft photons with energy $k \lsim 2T$,
and \ipa\ dominates the rate for very hard photons, $k \gg 10 \, T$.
For intermediate values of $k$, which is the most easily detected range
of photon energies, 
the $2 \leftrightarrow 2$ processes are of comparable importance to the
sum of the near-collinear inelastic processes.  
The LPM suppression is substantial precisely where the
inelastic mechanisms are dominant, but is fairly mild for intermediate values
of photon energy.
Parametrically, the soft photon emission rate behaves as
$d\Gamma_\gamma \propto k^{-1/2} \, dk$,
which is less infrared singular, by a square root of $k$,
from the result
found in Refs.~\cite{softgamma} and \cite{Gelis1} which neglect the LPM effect.


\begin{references}

\bibitem{oldie1}
E.~V.~Shuryak,
Phys.\ Lett.\ B {\bf 78}, 150 (1978)
[Sov.\ J.\ Nucl.\ Phys.\  {\bf 28}, 408.1978\ YAFIA,28,796 (1978)].

\bibitem{oldie2}
K.~Kajantie and H.~I.~Miettinen,
Z.\ Phys.\ C {\bf 9}, 341 (1981).

\bibitem{oldie3}
K.~Kajantie and P.~V.~Ruuskanen,
Phys.\ Lett.\ B {\bf 121}, 352 (1983).

\bibitem{oldie4}
F.~Halzen and H.~C.~Liu,
Phys.\ Rev.\ D {\bf 25}, 1842 (1982).

\bibitem{oldie5}
B.~Sinha,
Phys.\ Lett.\ B {\bf 128}, 91 (1983).

\bibitem{oldie6}
R.~C.~Hwa and K.~Kajantie,
Phys.\ Rev.\ D {\bf 32}, 1109 (1985).

\bibitem{oldie7}
G.~Staadt, W.~Greiner and J.~Rafelski,
Phys.\ Rev.\ D {\bf 33}, 66 (1986).

\bibitem{oldie8}
M.~Neubert,
Z.\ Phys.\ C {\bf 42}, 231 (1989).

\bibitem{Kapusta}
J.~Kapusta, P.~Lichard and D.~Seibert,
Phys.\ Rev.\ D {\bf 44}, 2774 (1991)
[Erratum--{\em ibid.}\ D~{\bf 47}, 4171 (1991)].

\bibitem{Baier}
R.~Baier, H.~Nakkagawa, A.~Niegawa and K.~Redlich,
Z.\ Phys.\ C {\bf 53}, 433 (1992).

\bibitem{Weldon}
H.~A.~Weldon,
Phys.\ Rev.\ D {\bf 26}, 2789 (1982).



\bibitem{Gelis1}
P.~Aurenche, F.~Gelis, R.~Kobes and H.~Zaraket,
Phys.\ Rev.\ D {\bf 58}, 085003 (1998)
[hep-ph/9804224].

\bibitem{Gelis2}
P.~Aurenche, F.~Gelis and H.~Zaraket,
Phys.\ Rev.\ D {\bf 61}, 116001 (2000)
[hep-ph/9911367].

\bibitem{Gelis3}
P.~Aurenche, F.~Gelis and H.~Zaraket,
Phys.\ Rev.\ D {\bf 62}, 096012 (2000)
[hep-ph/0003326].

\bibitem{SteffenThoma}
F.~D.~Steffen and M.~H.~Thoma,
Phys.\ Lett.\ B {\bf 510}, 98 (2001)
[hep-ph/0103044].

\bibitem{LP}
L.~D.~Landau and I.~Pomeranchuk,
Dokl.\ Akad.\ Nauk Ser.\ Fiz.\  {\bf 92} (1953) 535;
\\L.~D.~Landau and I.~Pomeranchuk,
Dokl.\ Akad.\ Nauk Ser.\ Fiz.\  {\bf 92} (1953) 735.

\bibitem{M1}
A.~B.~Migdal, Doklady Akad. Nauk S.~S.~S.~R.~{\bf 105}, 77 (1955).

\bibitem{M2}
A.~B.~Migdal,
Phys.\ Rev.\  {\bf 103}, 1811 (1956).

\bibitem{us}
P.~Arnold, G.~D.~Moore, and L.~G.~Yaffe,
``Photon Emission from Ultrarelativistic Plasmas,''
[hep-ph/0109064].

\bibitem{Zakharov}
B.~G.~Zakharov,
JETP Lett.\  {\bf 63}, 952 (1996)
[hep-ph/9607440];
{\em ibid.}
{\bf 65}, 615 (1997)
[hep-ph/9704255];
Phys.\ Atom.\ Nucl.\  {\bf 61} (1998) 838
[Yad.\ Fiz.\  {\bf 61} (1998) 924]
[hep-ph/9807540].

\bibitem{BDMPS}
R.~Baier, Y.~L.~Dokshitzer, A.~H.~Mueller and D.~Schiff,
Nucl.\ Phys.\ B {\bf 531} (1998) 403
[hep-ph/9804212].

\bibitem{LPM_QCD1}
R.~Baier, D.~Schiff and B.~G.~Zakharov,
Ann.\ Rev.\ Nucl.\ Part.\ Sci.\  {\bf 50}, 37 (2000)
[hep-ph/0002198].

\bibitem{LPM_QCD2}
B.~G.~Zakharov,
hep-ph/9807396;
JETP Lett.\  {\bf 73}, 49 (2001)
[Pisma Zh.\ Eksp.\ Teor.\ Fiz.\  {\bf 73}, 55 (2001)]
[hep-ph/0012360].

\bibitem{LPM_QCD3}
R.~Baier, Y.~L.~Dokshitzer, S.~Peigne and D.~Schiff,
Phys.\ Lett.\ B {\bf 345}, 277 (1995)
[hep-ph/9411409];
Phys.\ Rev.\ C {\bf 60}, 064902 (1999)
[hep-ph/9907267].

\bibitem{LPM_QCD4}
R.~Baier, Y.~L.~Dokshitzer, A.~H.~Mueller, S.~Peigne and D.~Schiff,
Nucl.\ Phys.\ B {\bf 483}, 291 (1997)
[hep-ph/9607355];
{\em ibid.}
{\bf 484}, 265 (1997)
[hep-ph/9608322].

\bibitem{softgamma}
P.~Aurenche, F.~Gelis, R.~Kobes and E.~Petitgirard,
Phys.\ Rev.\ D {\bf 54}, 5274 (1996)
[hep-ph/9604398].

\bibitem{LPM_QED1}
R.~Blankenbecler and S.~D.~Drell,
Phys.\ Rev.\ D {\bf 53}, 6265 (1996).

\bibitem{LPM_QED2}
R.~Baier, Y.~L.~Dokshitzer, A.~H.~Mueller, S.~Peigne and D.~Schiff,
Nucl.\ Phys.\ B {\bf 478}, 577 (1996)
[hep-ph/9604327].

\bibitem{LPM_QED3}
B.~G.~Zakharov,
Pisma Zh.\ Eksp.\ Teor.\ Fiz.\  {\bf 64}, 737 (1996)
[JETP Lett.\  {\bf 64}, 781 (1996)]
[hep-ph/9612431];
B.~G.~Zakharov,
Phys.\ Atom.\ Nucl.\  {\bf 62}, 1008 (1999)
[Yad.\ Fiz.\  {\bf 62}, 1075 (1999)]
[hep-ph/9805271];
{\em ibid.}
{\bf 61}, 838 (1998)
[hep-ph/9807540].

\bibitem{Weldon1}
H.~A.~Weldon,
Phys.\ Rev.\ D {\bf 26}, 1394 (1982).

\bibitem{lead_log}
P.~Arnold, G.~D.~Moore and L.~G.~Yaffe,
JHEP {\bf 0011}, 001 (2000)
[hep-ph/0010177].

\bibitem{largeN}
G.~D.~Moore,
JHEP {\bf 0105}, 039 (2001)
[hep-ph/0104121].

\end{references}
\end{document}